\documentclass[letterpaper]{article} 
\usepackage{main}  
\usepackage{times}  
\usepackage{helvet}  
\usepackage{courier}  
\usepackage[hyphens]{url}  
\usepackage{graphicx} 
\usepackage{natbib}  
\usepackage{caption} 
\usepackage[ruled,linesnumbered]{algorithm2e}
\usepackage{colortbl}
\usepackage{bbold}
\usepackage{multirow}
\usepackage{makecell}
\usepackage{pifont}
\usepackage{subfigure}
\usepackage{enumitem}
\usepackage{newtxmath}
\usepackage{booktabs}
\usepackage{xcolor}
\usepackage{subfiles}
\usepackage{amsmath}
\usepackage{graphicx}

\title{FINE: Factorized multimodal sentiment analysis via mutual INformation Estimation}
\author{
    Yadong Liu,
    Shangfei Wang\thanks{Corresponding author.}
}
\affiliations{
    University of Science and Technology of China \\
    yadongliu@mail.ustc.edu.cn, sfwang@ustc.edu.cn
}

\begin{document}

\maketitle

\begin{abstract}
Multimodal sentiment analysis remains a challenging task due to the inherent heterogeneity across modalities. Such heterogeneity often manifests as asynchronous signals, imbalanced information between modalities, and interference from task-irrelevant noise, hindering the learning of robust and accurate sentiment representations. To address these issues, we propose a factorized multimodal fusion framework that first disentangles each modality into shared and unique representations, and then suppresses task-irrelevant noise within both to retain only sentiment-critical representations. This fine-grained decomposition improves representation quality by reducing redundancy, prompting cross-modal complementarity, and isolating task-relevant sentiment cues. Rather than manipulating the feature space directly, we adopt a mutual information–based optimization strategy to guide the factorization process in a more stable and principled manner. To further support feature extraction and long-term temporal modeling, we introduce two auxiliary modules: a Mixture of Q-Formers, placed before factorization, which precedes the factorization and uses learnable queries to extract fine-grained affective features from multiple modalities, and a Dynamic Contrastive Queue, placed after factorization, which stores latest high-level representations for contrastive learning, enabling the model to capture long-range discriminative patterns and improve class-level separability. Extensive experiments on multiple public datasets demonstrate that our method consistently outperforms existing approaches, validating the effecti veness and robustness of the proposed framework.
\end{abstract}


\section{Introduction}

Sentiment analysis aims to uncover sentiments or opinions when individuals encounter specific topics, people, or entities~\cite{soleymani2017survey}. Since its inception, it has rapidly evolved into a vital research area with widespread applications in robotics, healthcare, education, and other industries~\cite{melville2009sentiment, petrovica2017emotion, liu2017facial, sanchez2019social}. As the internet transitions from text-based to a multimedia-driven space, sentiment analysis has dramatically transformed, leading to the rise of multimodal sentiment analysis (MSA). MSA integrates rich information from diverse modalities like text, audio, and images, providing a more comprehensive and accurate understanding of human sentiment~\cite{gandhi2023multimodal}. However, significant heterogeneity exists across modalities. Each modality not only reveals distinct types of sentiment cues  but also varies in representational density and noise. For example, text typically provides high semantic density~\cite{he2022masked}, while video and audio may contain redundant or overly complex sentiment representations.

\begin{figure}[tbp]
\centering
\includegraphics[width=0.47\textwidth]{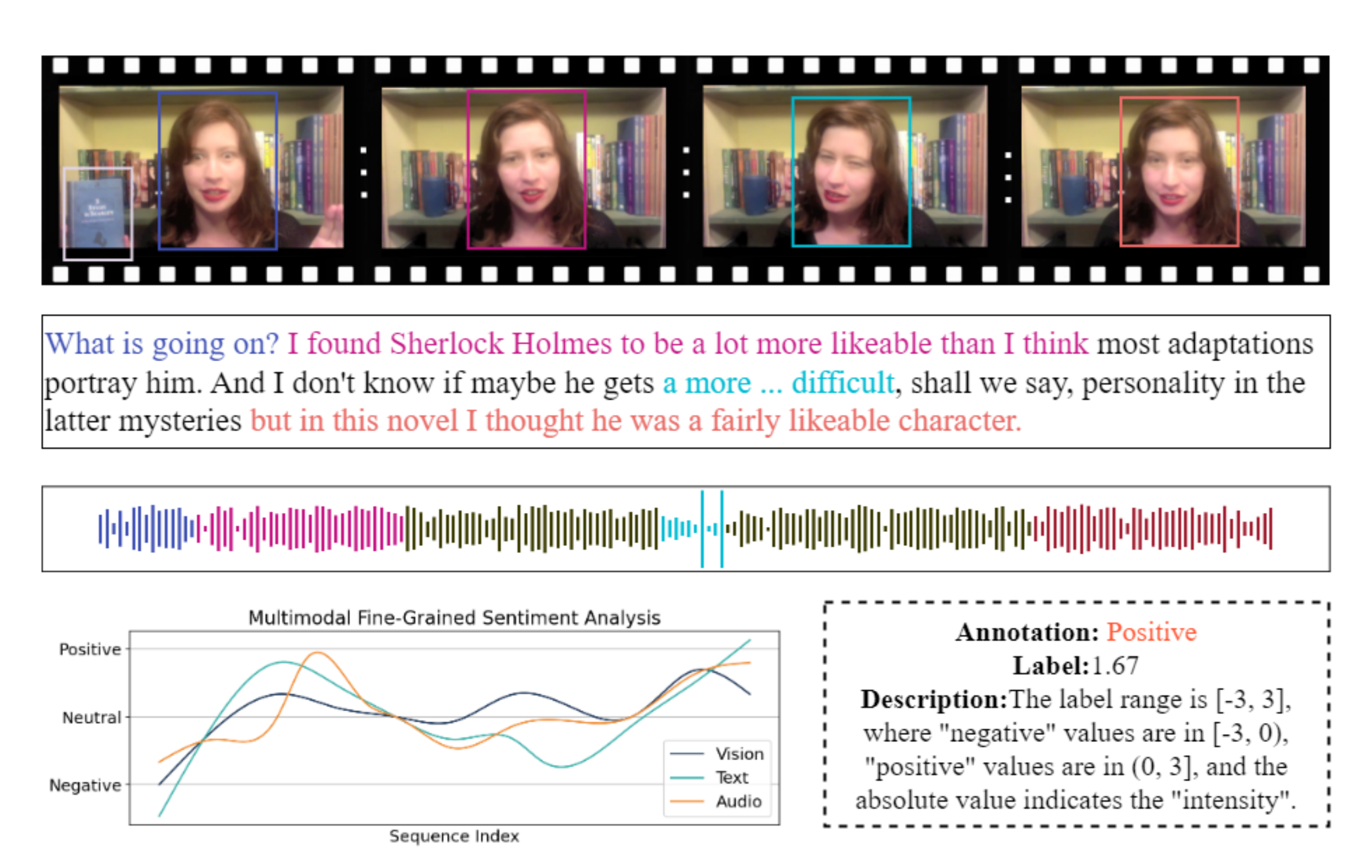}
\caption{A sample of MSA, incorporating three modalities: visual, textual, and audio. The bottom-left section displays fine-grained sentiment analysis, while the bottom-right section shows the label and annotation for this example.}
\label{motivation}
\end{figure}

MSA involves developing fusion strategies across different modalities to comprehensively predict sentiment polarity and intensity~\cite{morency2011towards}. Recent works primarily focus on multimodal representation learning and fusion, aiming to encode and integrate representations from diverse modalities to identify sentiment patterns. To achieve effective prediction, a range of approaches have been explored, including Multi-Layer Perceptrons (MLPs), attention-based models~\cite{sun2022cubemlp}, Long Short-Term Memory (LSTM) networks~\cite{tsai2018learning, lin2023dynamically}, Graph Neural Networks~\cite{mai2020modality}, and transformer encoders~\cite{tsai2019multimodal, rahman2020integrating, han2021bi, liang2022high}. Among these techniques, fusion strategies play a central role in aligning features from different modalities.

From the perspective of fusion strategies, MSA methods are commonly categorized into early, intermediate, and late fusion. Early fusion~\cite{liu2018efficient, poria2016convolutional} directly concatenates features from different modalities, but often fails to capture inter-modal heterogeneity. Late fusion~\cite{zhang2023provable} combines predictions from unimodal classifiers, yet lacks inter-modal interactions. Intermediate fusion~\cite{mai2020modality, nagrani2021attention, mai2022multimodal, ma2023multimodal, jiang2022sdn, fan2023pmr} offers a flexible trade-off, enabling joint embedding learning while reducing noise. Notably, models like Tensor Fusion Networks~\cite{zadeh2017tensor} and Multimodal Transformers~\cite{tsai2019multimodal} have demonstrated success in joint modeling. However, many of these methods treat each modality as a whole and perform fusion through simple operations such as concatenation or weighted averaging in a shared latent space. Such strategies often fail to capture modality-specific information and limit the model’s flexibility and interpretability in handling modality heterogeneity.

To more effectively address this issue, feature disentanglement has gained popularity. Representative works such as MISA~\cite{hazarika2020misa} project modality representations into two distinct subspaces. The first subspace captures modality-invariant features by aligning common semantics, while the second preserves modality-specific representations that reflect the private characteristics of each modality. DMD~\cite{li2023decoupled} introduces a graph distillation unit to dynamically decouple homogeneous and heterogeneous features. ConFEDE~\cite{yang2023confede} further decomposes modality features into similar and dissimilar features and employs contrastive loss to unify representation learning with disentanglement. However, these methods have yet to fully exploit the potential of disentanglement, as two subspaces may still contain task-irrelevant noise. Directly using them may result in optimization conflicts between the reconstruction and classification tasks. To address this, we design a two-stage disentanglement module. The first stage follows existing formulations by factorizing each modality into shared and unique representations and performing reconstruction. The second stage suppresses task-irrelevant noise in both branches, retaining only task-relevant representations for downstream sentiment modeling. Unlike previous approaches that directly manipulate the feature space using metrics such as Central Moment Discrepancy, Frobenius norm minimization, or contrastive loss, our method adopts a mutual information estimation objective, providing a more stable, information-theoretic foundation for disentanglement.


 As illustrated in Figure~\ref{motivation}, multimodal sentiment cues are often temporally heterogeneous. In the visualization, we align content from different modalities using consistent color coding—segments with the same color correspond to the same time step. At the beginning, the segment \textit{``What is going on?''} expresses mild confusion in both audio and text, but it deviates from the final sentiment label and acts as task-irrelevant noise. The phrase \textit{``I found Sherlock Holmes to be a lot more likeable...''} conveys consistent positivity across all three modalities, reflecting shared task-relevant cues. However, the facial expression and intonation corresponding to this sentiment vary in timing and intensity, indicating asynchronous signals and imbalanced information across modalities. Later, in \textit{``a more... difficult,''} sentiment becomes ambiguous: the text implies negativity, while the speaker’s playful expression and tone suggest otherwise. This highlights inconsistency in sentiment across modalities and suggests the presence of unique information in each channel. Only by integrating all modalities can this subtle positivity be accurately interpreted, emphasizing the importance of cross-modal complementarity. These observations suggest that multimodal sentiment analysis faces three key challenges: asynchronous signals, imbalanced information, and interference from task-irrelevant noise. These issues stem from inherent modality heterogeneity, which hinders the learning of robust and accurate sentiment representations. Therefore, there is a critical need for a principled fusion strategy that can disentangle shared and unique features, suppress task-irrelevant noise, and leverage cross-modal synergy to reduce redundancy, enhance complementarity, and improve sentiment modeling.

To overcome these issues, we propose FINE, a factorized multimodal sentiment analysis framework grounded in mutual information estimation. FINE first employs a Mixture of Q-Formers to extract fine-grained sentiment cues from each modality at an early stage using learnable queries. These extracted features are then fed into the Factorized Task-Relevant Encoder, which factorizes each extracted feature into shared and unique representations, and further removes task-irrelevant noise from both by optimizing mutual information objectives. A Transformer encoder is subsequently used to fuse the task-relevant features across modalities, integrating semantic and affective cues. To support long-term pattern modeling and improve representation robustness, we introduce a Dynamic Contrastive Queue that stores recent high-level features for temporal contrastive learning. Together, these components enable FINE to effectively model complex, noisy, and asynchronous multimodal sentiment signals. Our contributions are summarized as follows:
\begin{itemize}
    \item We propose a factorized framework that jointly disentangles shared/unique and task-relevant/irrelevant features, enhancing robustness and interpretability.
    \item We introduce a query-based extraction mechanism for fine-grained sentiment representations and a contrastive learning module for capturing long-range dependencies.
    \item Experiments on multiple benchmarks validate the effectiveness of FINE and demonstrate the utility of each component.
\end{itemize}

\section{Methodology}
\begin{figure}[tbp]
    \centering
    \includegraphics[width=0.98\columnwidth]{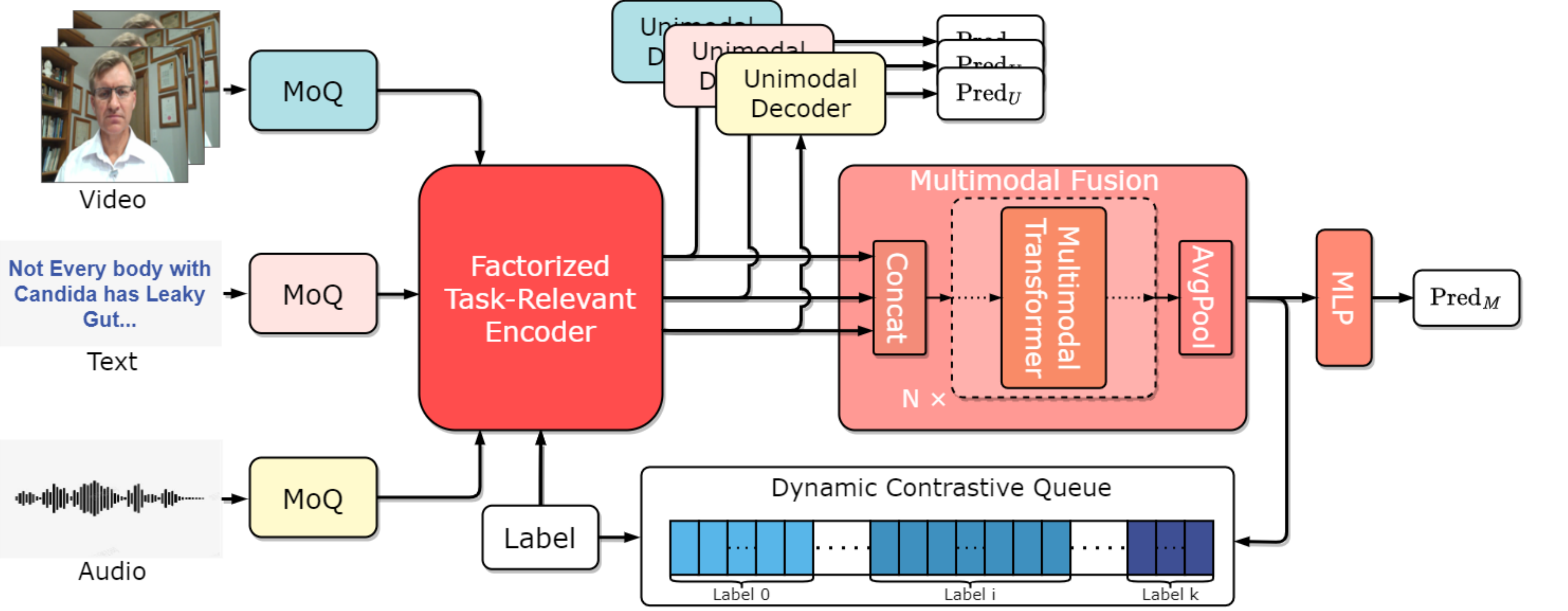}
    \caption{The overview of FINE. \(\mathrm{Pred}_{U}\) represents the unimodal prediction, while \(\mathrm{Pred}_{M}\) corresponds to the multimodal prediction.}
    \label{main}
\end{figure}

As illustrated in Figure~\ref{main}, FINE first processes raw modality features using a set of Q-Former experts (MoQ). Each expert employs $L$ learnable query tokens to extract temporally aligned, fine-grained sentiment cues from the original multimodal sequences, effectively mitigating cross-modal temporal inconsistencies. The weighted query results from each modality are then passed to the Factorized Task-Relevant Encoder, which disentangles them into shared and unique task-relevant features. These six types of features (two from each modality) are concatenated and fed into a Transformer encoder to capture high-level semantic and sentiment dependencies across modalities. Finally, FINE introduces a Dynamic Contrastive Queue, which stores the latest task-relevant representations in class-aware sub-queues, promoting better discrimination across classes and preserving meaningful distance structures that reflect class-level distinctions. The following subsections detail the structure and functionality of each module.

\subsection{Mixture of Q-Formers}

To effectively capture modality-specific and fine-grained sentiment cues, we integrate a Mixture-of-Experts (MoE) architecture into our framework. MoQ is a straightforward extension of MoE for multimodal sentiment analysis, where each expert is designed to model diverse input subspaces across modalities. Instead of using conventional feedforward networks (FFNs) as experts, we employ Q-Formers~\cite{li2023blip}, which act as information bottlenecks by leveraging learnable query tokens to extract condensed sentiment representations. This design enables MoQ to retain expressive capacity for nuanced sentiment features while significantly reducing computational overhead. To ensure routing flexibility and prevent the model from collapsing into static expert usage, we adopt an auxiliary load-balancing loss \( \mathcal{L}_{aux} \), which encourages more uniform expert activation during training.

Formally, for a modality feature \( X_m \in \mathbb{R}^{T_m \times d_m} \), where \( m \in \{T, A, V\} \) denotes text, audio, or visual modality, the MoQ module produces a compressed output of \( l \) tokens:
\begin{equation}
\hat{x}_m^i = \text{MoQ}_m(x_m^i),
\end{equation}
where \( \hat{x}_m^i \in \mathbb{R}^{l \times d_m'} \), and \( d_m' \) denotes the hidden dimensionality of the Q-Former outputs.

Further details on expert routing and the internal structure of Q-Formers are provided in the Supplementary Material.

\subsection{Factorized Task-Relevant Encoder}

The Factorized Task-Relevant Encoder (FTRE) is designed as a general-purpose module applicable to a broad range of tri-modal supervised tasks in real-world scenarios. Given a input space consisting of three modalities \(X_1\), \(X_2\), and \(X_3\), FTRE assumes that the task-relevant information can be factorized into two types: shared information and unique information. The former denotes information that is common across multiple modalities, while the latter captures information specific to individual modalities. Both types of information are essential for accurately modeling the target variable \(Y\).

\begin{figure}[!t]
    \centering
    \includegraphics[width=0.46\textwidth]{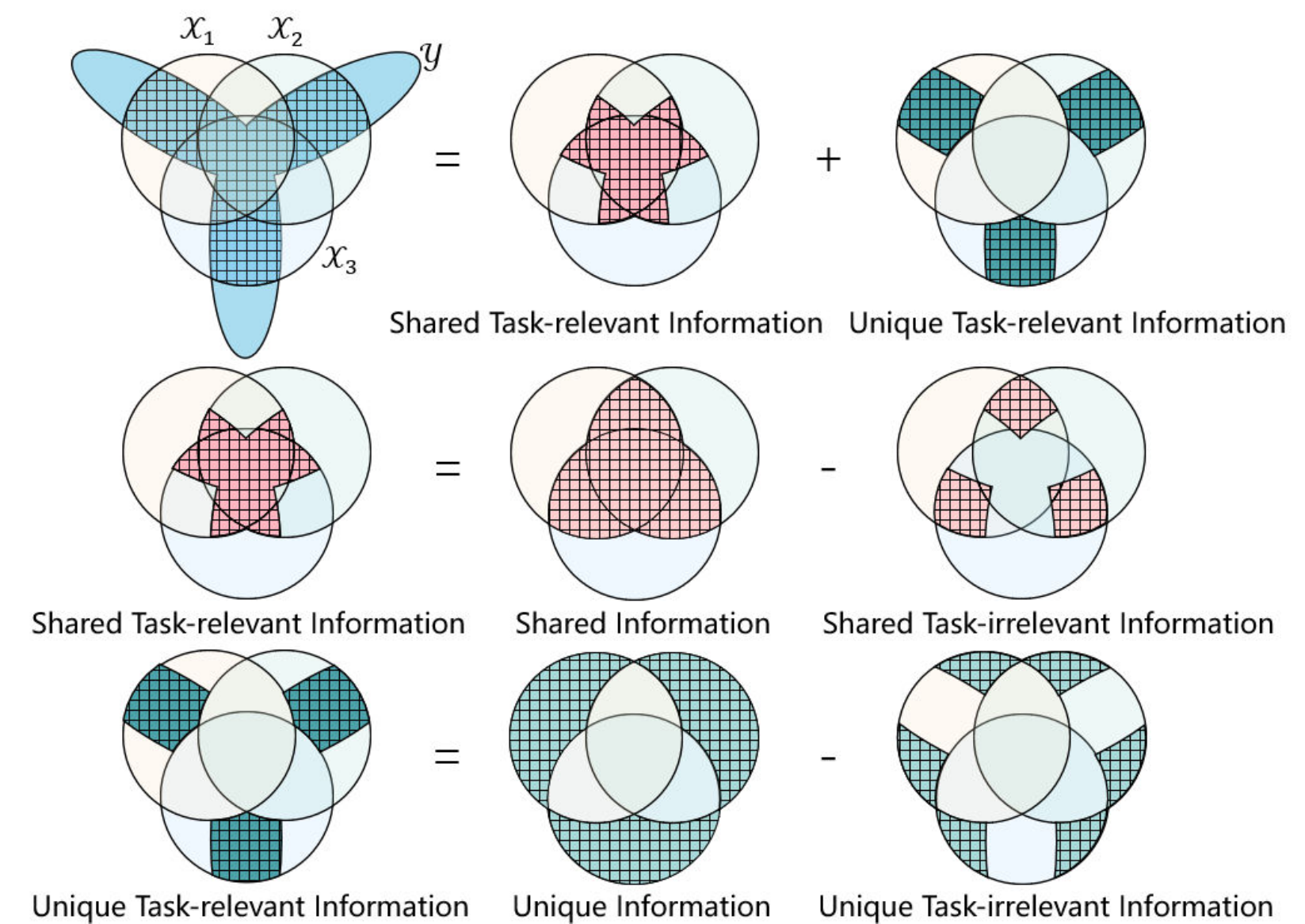}
    \caption{A Venn diagram illustrating mutual information among modalities $\mathcal{X}_1$, $\mathcal{X}_2$, $\mathcal{X}_3$ and the task $\mathcal{Y}$. The blue region denotes task-relevant information, with the blue grid marking the ideal target for learning. This target combines task-relevant shared information (red) and unique information (green). Achieving it requires suppressing task-irrelevant noise from both components.
 }
    \label{FTREvn}
    \vspace{-0.3cm}
\end{figure}

As demonstrated in the Venn diagram in Figure~\ref{FTREvn}, the ideal task-relevant information is factorized into two conditional mutual information terms in the tri-modal feature space: one representing the shared task-relevant information \(S_{TR}\), and the other representing the unique task-relevant information \(U_{TR}\). This factorization is achieved in two steps, as follows:

\begin{equation}
    I(X_1, X_2, X_3; Y) = S_{TR} + U_{TR}=(S - S_{TI}) +(U - U_{TI}) \, , \label{eq:mi}
\end{equation}
where \(S\) represents the shared information between any two modalities \(X_i\) and \(X_j\), and \(U\) represents the unique information specific to each modality. Specifically, \(S\) can be expressed as the aggregation of pairwise mutual information across modalities. On the other hand, \(U\) can be expressed as the sum of the conditional mutual information between one modality and the rest:

\begin{equation}
    S = \sum_{1 \leq i < j \leq 3} I(X_i; X_j) \, , 
    U = \sum_{
        \substack{
            \{i, j, k\} \subseteq \{1, 2, 3\} \\
            i \neq j \neq k
        }
    } I(X_i \mid X_j, X_k) \, ,
\end{equation}

and \(S_{TR}\) represents the mutual information corresponding to the shared task-relevant information. In contrast, \(S_{TI}\), which is irrelevant to the task, represents noisy information that should be excluded. \(S_{TI}\) can be defined as the union of pairwise mutual information conditioned on the task \(Y\). Directly solving for \(S_{TR}\) is not very practical, so we indirectly approximate the  \(S_{TR}\) by minimizing \(S_{TI}\):
\begin{equation}
S_{TI} = \sum_{1 \leq i < j \leq 3} I(X_i; X_j \mid Y) \, ,
\end{equation}
and \(U_{TR}\) denotes the unique task-relevant information. This term highlights the distinct contributions of each modality to the task and can be expressed as the union of the mutual information between each modality and the task:  
\begin{equation}
U_{TR} = \sum_{1 \leq i \leq 3} I(X_i;Y) \, .
\end{equation}

\begin{figure}
    \centering
    \includegraphics[width=0.45\textwidth]{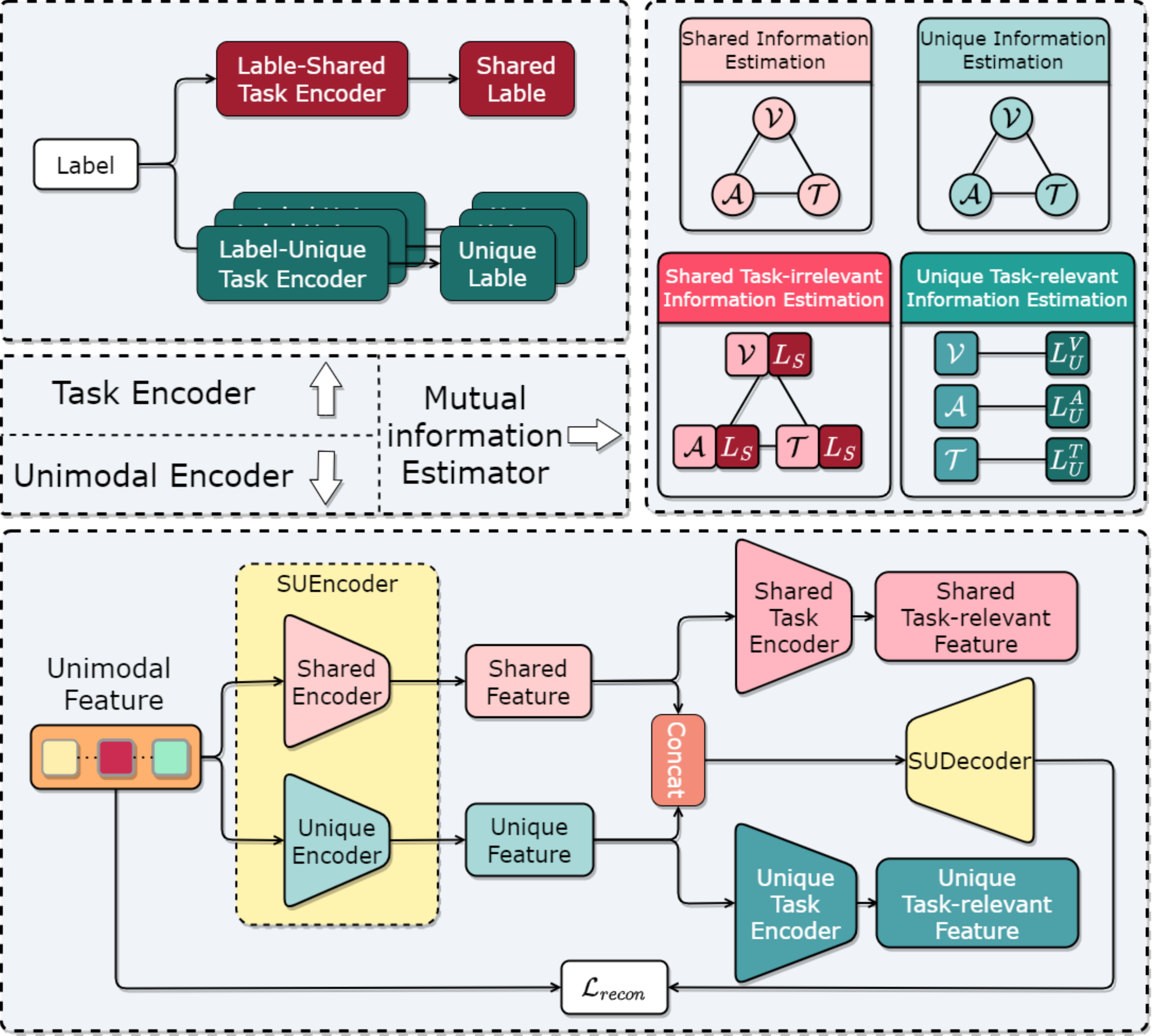}
    \caption{The structure of FTRE. The bottom-left corner depicts the unimodal encoders, where each modality \(X_i\) is processed independently. The top-left corner illustrates the direct encoding of labels, resulting in one shared and three unique label embeddings. The top-right corner represents the estimation of four distinct types of mutual information.}
    \label{FTRE}
\end{figure}

In the following section, we will transition from the theoretical analysis presented above to the practical implementation of the FTRE module for the extraction of \(S_{TR}\) and \(U_{TR}\). The detailed architecture of FTRE is shown in Figure~\ref{FTRE}. Specifically, within the Unimodal Encoder, we propose the Shared\&Unique Encoder (SUEncoder), which projects the input modality features \(X_m\) into two distinct representations: one representing the shared information across modalities, denoted as \(S_m\), and the other capturing the unique information, denoted as \(U_m\). The SUEncoder consists of MLPs with activation functions, referred to as the shared encoder (\(E_m^s\)) and the unique encoder (\(E_m^u\)).

\begin{equation}
x_{m, i}^s = E_m^{s}(\hat{x}_m^i) \, ,\quad x_{m,i}^u = E_m^{u}(\hat{x}_m^i)  \, .
\end{equation}

To compute \(S\) in Equation~\eqref{eq:mi}, we introduce the InfoNCE~\cite{oord2018representation} to maximize a lower bound on the mutual information between shared features \(x_{m,i}^s\) from any two modalities,

\begin{equation}
    I_{sha}\left(X_1^s ; X_2^s\right)=
    \mathbb{E}_{\substack{x_1, x_2^+ \sim p\left(x_1, x_2\right) \\ x_{2}^{-} \sim p\left(x_2\right)}}\left[\log \frac{\exp f\left(x_1, x_2^+\right)}{\sum_{k} \exp f(x_1, x_2^-)}\right]  \, ,
\end{equation}
Where \( f(x_1, x_2^+) \) is the optimal critic, and \( x_2^+ \) refers to the feature of another modality from the same sample as \( x_1 \), while \( x_2^- \) denotes the feature from a different sample.

To compute \(U\) in Equation~\ref{eq:mi}, the NCE-CLUB~\cite{liang2024factorized} is introduced, which minimizes the expected upper bound on the unique feature \(x_{m,i}^u\) between any two modalities. It effectively achieves this “for free” while avoiding the need to separately optimize both the lower bound and upper bound:
\begin{equation}
    \begin{aligned}
        I_{uni}(X_1^u ; X_2^u) &= 
        \, \mathbb{E}_{x_1, x_2^+ \sim p(x_1, x_2)}\left[f^{\ast}(x_1,x_2^+)\right] 
        \\ & - \, \mathbb{E}_{\substack{x_1 \sim p(x_1) \\ x_2^- \sim p(x_2)}}\left[f^{\ast}(x_1, x_2^-) \right],
    \end{aligned}
\end{equation}
where \( f^{\ast}(x_1, x_2^+) \) is the optimal critic from \( I_{NCE} \), used within the \( I_{CLUB} \)~\cite{cheng2020club}.

The aforementioned operations establish a disentanglement mechanism, ensuring that both representations retain as much of information from the original modality as possible. To reinforce information preservation, we incorporate a decoder that reconstructs the original modality features \( \hat{X}_m \)  from the representations of the shared and unique information. The reconstruction process is guided by a reconstruction loss \(\mathcal{L}_{recon}\), calculated using Mean Squared Error, which encourages accurate recovery of the modality-specific input. To further suppress task-irrelevant noise, we feed the unique and shared features, \( x_{m,i}^u \) and \( x_{m,i}^s \), into their respective task encoders, \( E_m^{utr} \) and \( E_m^{str} \). Both \( E_m^{utr} \) and \( E_m^{str} \) are implemented as MLPs, which serve to refine the representations by filtering out noise while preserving task-relevant semantics, as formulated in the following equations:
\begin{equation}
x_{m, i}^{str} = E_m^{str}(x_{m, i}^s) \, ,\quad x_{m,i}^{utr} = E_m^{utr}(x_{m, i}^u)  \, .
\end{equation}

To ensure the effective realization of the aforementioned functionality, we introduce the conditional InfoNCE estimator and the NCE-CLUB estimator as constraints to guarantee that the extracted information remains task-relevant. Specifically, for the concatenated feature set of \( X_m^{str} \in \mathbb{R}^{d} \) from modality \( m \) and \( Y^{str} \),  where \( Y^{str} \in \mathbb{R}^{d_s}\) is obtained by passing the original label \( Y \) through a Label-Shared Task Encoder, the mutual information between this feature set and the concatenated feature set of \( X_n^{str} \) from another modality \( n \) along with \( Y^{str} \) is minimized in its upper bound using NCE-CLUB. This procedure effectively minimizes the task-irrelevant mutual information:

\begin{align}
I_{str}(X_1^{str} ; X_2^{str} \mid Y^{str}) 
&= \mathbb{E}_{p(y)} \Big[ \notag \\
& \quad \mathbb{E}_{x_1, x_2^+ \sim p(x_1, x_2 \mid y)} \left[f^*(x_1, x_2^+, y)\right]\notag \\
&\quad - \mathbb{E}_{\substack{x_1 \sim p(x_1 \mid y) \\ x_2^- \sim p(x_2 \mid y)}} \left[f^*(x_1, x_2^-, y) \right]
\Big].
\end{align}


Similarly, we employ the InfoNCE estimator to maximize the lower bound of the mutual information between \( X_m^{utr} \in \mathbb{R}^{d} \) from modality \( m \) and \( Y_m^{utr} \in \mathbb{R}^{d_u} \), where \( Y_m^{utr} \) is obtained by passing the original label \( Y \) through a Label-Unique Task Encoder. This operation serves to reduce the task-irrelevant noise present within the unique features:
\begin{equation}
    I_{utr}\left(X_1^{utr} ; Y_1^{utr}\right)=
    \mathbb{E}_{\substack{x_1, y_1^+ \sim p(x_1, y_1) \\ y_1^{-} \sim p(y_1)}}\left[\log \frac{\exp f\left(x_1, y_1^+\right)}{\sum_{k} \exp f(x_1, y_1^-)}\right] ,
\end{equation}
where \( y_1^+ \) represents a positive label related to \( x_1 \), while \( y_1^- \) is a negative label from the marginal distribution \( p(y) \).

The total mutual information loss integrates the four mutual information estimations discussed above. Specifically, $I_{sha}$ and $I_{utr}$ are estimated as lower bounds and optimized by minimizing their negatives, while $I_{uni}$ and $I_{str}$ are treated as upper bounds and directly minimized. The overall mutual information loss \(\mathcal{L}_{MI}\) is given by:

\begin{equation}
\mathcal{L}_{MI} = - I_{sha} + I_{uni} + \mathcal{L}_{recon}+ I_{str} -  I_{utr}.
\end{equation}

The final output of each modality is obtained by concatenating $X_m^{str}$ and $X_m^{utr}$, thereby integrating task-relevant information extracted from both shared and unique features.

\subsection{Dynamic Contrastive Queue}
To capture long-term discriminative patterns and mitigate incidental noise, we incorporate a Dynamic Contrastive Queue strategy. This queue-based mechanism uses class-wise sub-queues to retain recent task-relevant features over time. It allows the model to encourage greater inter-class representation separation, effectively addressing class imbalance and noise sensitivity. Additionally, we adopt the Angle-Compensated Contrastive Regularizer (ACCon)~\cite{zhao2025accon} to adjust similarity based on label differences, improving representation precision in continuous sentiment spaces. Finally, at training step \(t\), we compute the angle-compensated contrastive loss \(\mathcal{L}^{i}_{CL}\) for sample \(i\) using the updated queue:
\begin{equation}
    \mathcal{L}^{i}_{CL} = \frac{-1}{|\mathcal{P}(i)|}
    \sum_{p \in \mathcal{P}(i)} 
    \log \frac{\exp (z_{i} z_{p}^{T} / \tau)}
    {\left( 
    \substack{
            \sum_{k \in \mathcal{P}(i)} \exp (z_{i} z_{k}^{T} / \tau ) \\
            +\sum_{m \in \mathcal{N}(i)} \exp (\cos (\tilde{\theta}_{i, m}) / \tau)
    }
    \right)} \, ,
\end{equation}
where \( \mathcal{P}(i) \) and \( \mathcal{N}(i) \) represent the positive and negative sample sets for anchor \( i \), sampled from both the current batch at step \(t\) and the historical queue accumulated over the previous \(t-1\) steps. Further details, including the queue update process and angle-aware similarity compensation, are provided in the Supplementary Material.

\subsection{Multimodal Fusion and Prediction}
We combine the shared task-relevant features and the unique task-relevant features to construct multimodal representations for downstream tasks, which include both auxiliary unimodal prediction and multimodal fusion-based prediction. To extract and fuse these multimodal features, we leverage the Transformer ~\cite{vaswani2017attention} to learn comprehensive multimodal sentiment knowledge. Additionally, we use the [CLS] token of each each modality as a representation of the information of that modality. Then, fully connected layers are applied to predict the final results based on these fused multimodal representations.

We integrate the above losses to formulate the comprehensive optimization objective:
\begin{equation}
\begin{aligned}
    \mathcal{L}_{total}= & \mathcal{L}_{MP} + \lambda_{up} \mathcal{L}_{UP} + \lambda_{cl}\mathcal{L}_{CL} \\ 
    & + \lambda_{aux}\mathcal{L}_{aux} +\beta_{mi} \mathcal{L}_{MI} \, ,
\end{aligned}
\end{equation}
where \(\mathcal{L}_{MP}\) and \(\mathcal{L}_{UP}\) represent the multimodal and unimodal prediction losses, respectively, defined as the Mean Squared Error in our experiments. The hyperparameters \(\lambda_{cl}\), \(\lambda_{up}\), \(\lambda_{aux}\), and \(\beta_{mi}\) control the relative contribution of the different loss components to the overall optimization.

\section{Experiments}
\subsection{Experimental Settings}

\newcommand{\mysubsubsection}[1]{\subsubsection*{\noindent \normalfont \textbf{#1}}}

\mysubsubsection{Dataset}
To validate the effectiveness of the proposed model, we conducted experiments on four widely-used multimodal datasets: CMU-MOSI~\cite{zadeh2016mosi}, CMU-MOSEI~\cite{zadeh2018multimodal}, UR-FUNNY~\cite{hasan2019ur}, and CH-SIMS~\cite{yu2020ch}. These datasets span a variety of domains and languages, enabling a comprehensive evaluation of model performance under different multimodal sentiment analysis scenarios. Detailed descriptions of the datasets, evaluation metrics, implementation settings, and baseline models can be found in the supplementary material.

\begin{table}[b]
\centering
{\fontsize{9}{10}\selectfont
\setlength{\tabcolsep}{1mm}
\begin{tabular}{c*{3}{c}}
\toprule
 \textbf{Model} & \textbf{ACC-2 (↑)} & \textbf{F1 (↑)} & \textbf{ACC-7 (↑)} \\
\midrule
TFN~\cite{zadeh2017tensor} & 80.8  & 80.7  & 34.9 \\
LMF~\cite{liu2018efficient}  & 82.5  & 82.4  & 33.2 \\
GFN~\shortcite{mai2020modality} & 84.3  & 84.3  & 47.0  \\
MulT~\cite{tsai2019multimodal} & 83.7  & 83.7  & 41.5  \\
CubeMLP~\cite{sun2022cubemlp} & 85.6  & 85.5  & 45.5 \\
MISA & 83.4  & 83.6  & 42.3 \\
BBFN~\cite{han2021bi} & 84.3  & 84.3  & 45.0 \\
C-MIB~\shortcite{mai2022multimodal} & 85.2  & 85.2  & \underline{48.2}  \\
MSG~\cite{lin2023dynamically} & 85.7  & 85.6  & 45.3 \\
ConFEDE~\cite{yang2023confede} & 85.52 & 85.52 & 42.27 \\
DMD~\shortcite{li2023decoupled} & 86.0 & 86.0  & 45.6 \\
EUAR~\cite{gao2024enhanced} & \underline{86.3} & \underline{86.3} & 46.1 \\
\midrule
\textbf{FINE (Ours)} & \textbf{86.95} &\textbf{86.94} &\textbf{48.50} \\
\bottomrule
\end{tabular}
}
\caption{Results on CMU-MOSI dataset.}
\label{performance_comparison_mosi}
\end{table}

\begin{table}[t]
\centering
{\fontsize{9}{10}\selectfont
\setlength{\tabcolsep}{1mm}
\begin{tabular}{c*{3}{c}}
\toprule
\textbf{Model} & \textbf{ACC-2 (↑)} & \textbf{F1 (↑)} & \textbf{ACC-7 (↑)} \\
\midrule
TFN~\cite{zadeh2017tensor} & 82.5  & 82.1  & 50.2  \\
LMF~\cite{liu2018efficient}  &  82.0  & 82.1  & 48.0  \\
GFN~\shortcite{mai2020modality} & 85.0  & 85.0  & 51.8  \\
MulT~\cite{tsai2019multimodal} & 84.7  & 84.6  & 50.7  \\
CubeMLP~\cite{sun2022cubemlp} & 85.1  & 84.5  & 54.9  \\
MISA~\shortcite{hazarika2020misa} & 85.5  & 85.3  & 52.2  \\
BBFN~\cite{han2021bi} & 86.2  & 86.1  & 54.8  \\
C-MIB~\shortcite{mai2022multimodal} & 86.2  & 86.2  & 53.0  \\
MSG~\cite{lin2023dynamically} & 85.4  & 85.4  & 52.8  \\
ConFEDE~\cite{yang2023confede} &  85.82 & 85.83 & 54.86 \\
DMD~\shortcite{li2023decoupled} & \underline{86.6}  & \underline{86.6}  & 54.5  \\
EUAR~\cite{gao2024enhanced} & \underline{86.6}  & 86.4  & \underline{54.9}  \\
\midrule
\textbf{FINE (Ours)} & \textbf{87.70} & \textbf{87.68} &\textbf{55.59}  \\
\bottomrule
\end{tabular}
}
\caption{Results on CMU-MOSEI datasets.}
\label{performance_comparison_mosei}
\end{table}

\begin{table}[t]
\centering
{\fontsize{9}{10}\selectfont
\setlength{\tabcolsep}{1mm}
\begin{tabular}{*{3}{c}|c}
\toprule
\textbf{Model} & Context & Target & \textbf{ACC-2 (↑)} \\
\midrule

C-MFN~\cite{hasan2019ur} & \ding{51} & \ding{51} & 65.23 \\
LMF~\cite{liu2018efficient} &  & \ding{51} & 67.53 \\
TFN~\cite{zadeh2017tensor} &  & \ding{51} & 68.57 \\
MISA~\shortcite{hazarika2020misa} &  & \ding{51} & 70.61 \\
MAGBERT(XLNet)~\shortcite{rahman2020integrating} &  & \ding{51} & 72.43 \\
MuLOT~\shortcite{pramanick2022multimodal} &  & \ding{51} & 73.22 \\
MuLOT~\shortcite{pramanick2022multimodal} & \ding{51} & \ding{51} & 
\underline{73.97} \\
\midrule
\textbf{FINE (Ours)} &  & \ding{51} &\textbf{74.95}  \\
\bottomrule
\end{tabular}
}
\caption{Results on UR-FUNNY datasets.}
\label{performance_comparison_urfunny}
\end{table}

\begin{table}[t]
\centering
{\fontsize{9}{10}\selectfont
\setlength{\tabcolsep}{1mm}
\begin{tabular}{c*{3}{c}}
\toprule
\textbf{Model} & \textbf{ACC-2 (↑)} & \textbf{F1 (↑)} & \textbf{ACC-5 (↑)} \\
\midrule
LF-DNN~\cite{yu2020ch} & 78.87 & 79.87  & 41.62  \\
MFN(A)~\cite{zadeh2018memory} & 78.87 & 79.87  & 39.47  \\
LMF~\cite{liu2018efficient}  &  77.77 & 77.88  & 40.53  \\
TFN~\cite{zadeh2017tensor} & 78.38 & 78.62  & 39.30  \\
MulT(A)~\cite{tsai2019multimodal} & 78.38 & 78.62  & 37.94  \\
Self-MM~\cite{yu2021learning} & 80.04 & 80.44 & 41.53  \\
ConFEDE~\cite{yang2023confede} &  82.23 & 82.08 & \textbf{46.30} \\
Coupled Mamba~\shortcite{li2024coupled} &  81.8 & 81.3 & \underline{42.1} \\
\midrule
\textbf{FINE (Ours)} & \textbf{82.28} & \textbf{82.22} & 41.79  \\
\bottomrule
\end{tabular}
}
\caption{Results on CH-SIMS datasets. (A) means the model utilized the aligned data.}
\label{performance_comparison_chsims}
\end{table}

\subsection{Results and Analysis}

\mysubsubsection{Quantitative Results}Table~\ref{performance_comparison_mosei} and Table~\ref{performance_comparison_mosi} present a comparative analysis of FINE with recent state-of-the-art models on the CMU-MOSEI and CMU-MOSI datasets. The best results are highlighted in bold font, and the second-best results are underlined. As shown in the table, on CMU-MOSEI, FINE yields consistent gains across all metrics, with 1-point improvements on ACC-2 and F1, and a 0.7-percentage-point gain on ACC-7. On CMU-MOSI, FINE sets new benchmarks in ACC-2, F1, and ACC-7. We hypothesize that the limited size of the CMU-MOSI dataset may have limited the expressiveness of FINE.  Table~\ref{performance_comparison_urfunny} evaluates FINE on the multimodal humor recognition dataset UR-FUNNY. FINE achieves a new state-of-the-art ACC-2 score of 74.95\%, outperforming the best previous model MAGBERT by 2.5 points. In Table~\ref{performance_comparison_chsims}, we also report results on CH-SIMS, a challenging Chinese-language sentiment benchmark with real-world video data. FINE achieves the best results on both ACC-2 and F1, and ranks second on ACC-5. These results demonstrate the effectiveness of our method in cross-lingual and more diverse multimodal settings. These improvements can be attributed to the disentanglement mechanism in FINE, which explicitly separates task-relevant shared and unique information while suppressing noise. Furthermore, MoQ facilitates early-stage alignment of multimodal representations, and DCQ improves long-range dependency modeling and class-level discrimination. Together, these design choices allow FINE to learn robust and complementary representations, leading to its superior performance across diverse benchmark settings.

\begin{figure}[htbp]
    \centering
    \subfigure[Classification]{
    \includegraphics[width=0.33\linewidth]{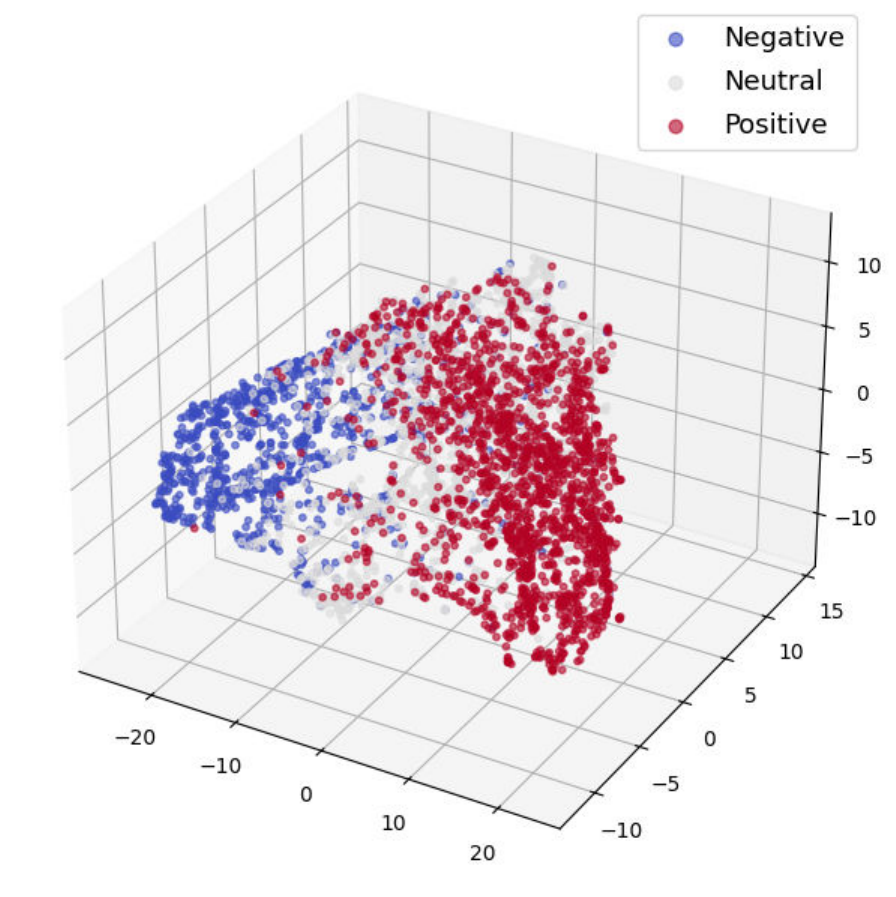}
    \label{fig:Visualizationfc}}
    \hspace{5pt}
    \subfigure[Different Modalities]{
    \includegraphics[width=0.52\linewidth]{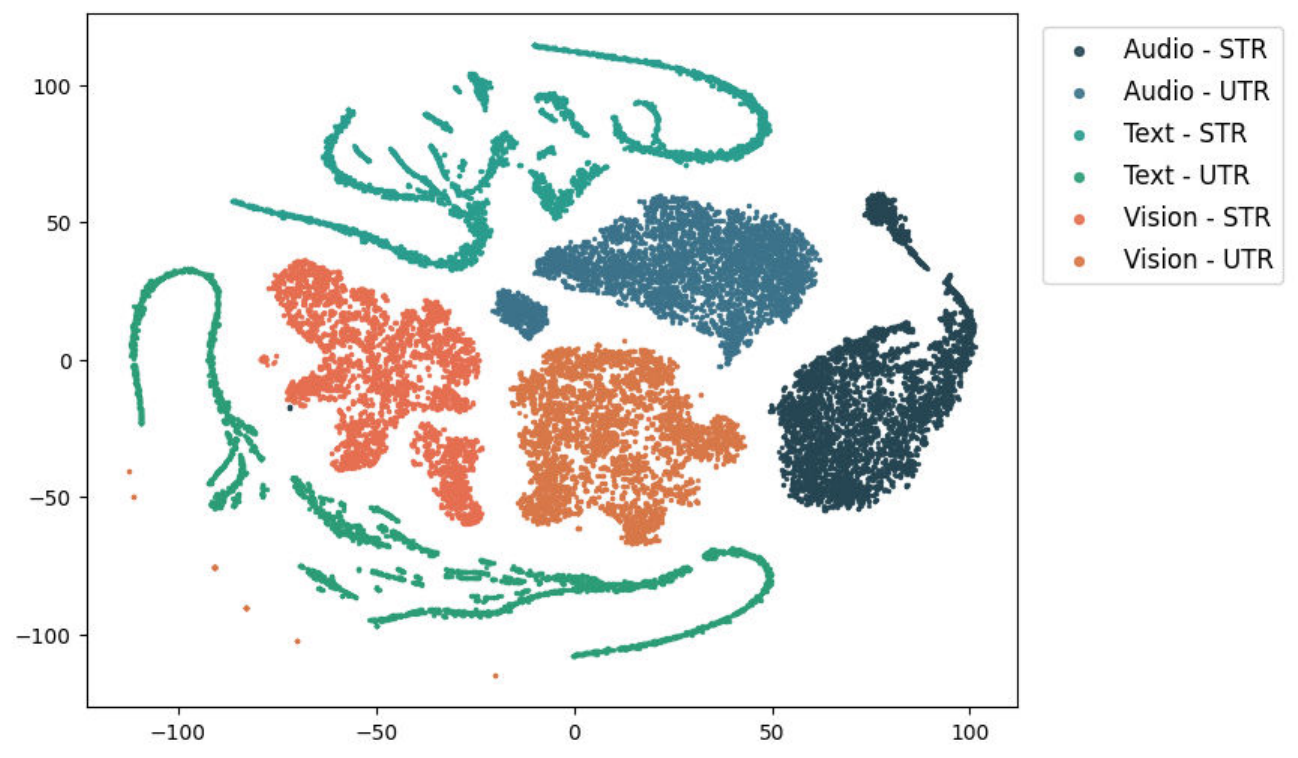}
    \label{fig:Visualizationfdm}}
    \caption{Visualization of features obtained after FTRE.}
    \label{tsne_after_ftre}
\end{figure}

\mysubsubsection{Visualization of Representations} Additionally, we visualized the representations obtained from the FTRE module, which fuses all three modalities. Each modality’s features were further decomposed into Shared Task-Relevant (STR) and Unique Task-Relevant (UTR) features. These six sets of features were concatenated and then used for classification visualization. The results are shown in Figure~\ref{tsne_after_ftre}, based on the test set of the CMU-MOSEI dataset. In Figure~\ref{fig:Visualizationfc}, we present the 3D visualization between learned features and sentiment labels. Blue, red, and gray represent Negative, Positive, and Neutral classes, respectively. From the density of color distributions and their spatial layout, it is evident that the representations learned by the FTRE module are strongly correlated with sentiment polarity. Moreover, Neutral points are mostly located between the Negative and Positive clusters, which aligns well with human sentiment intuition.

Figure~\ref{fig:Visualizationfdm} further illustrates that the six types of features exhibit both strong separability and cross-modal complementarity. Notably, STR and UTR features from the same modality display similar distributions, while those from different modalities are distinctly separated. This suggests that our method preserves the structural characteristics of each modality without enforcing strong constraints in the feature space. Furthermore, all six feature types show relatively high internal density, indicating that they capture high-level sentiment semantics while effectively filtering out irrelevant noise.

\subsection{Ablation Study}
In this section, we analyze the impact of each modality on performance using the CMU-MOSEI dataset and conduct ablation studies on key components of FINE. To assess their importance, we replace each with functionally similar alternatives and report results in Table~\ref{tab:ablation_for_modal} and Table~\ref{tab:ablation_for_modules}.

\begin{table}
    \centering
        \begin{tabular}{*{3}{c}|*{3}{c}}
            \hline
            \multicolumn{3}{c}{\textbf{Modality}} & \multicolumn{3}{c}{\textbf{Metrics}} \\
            \cmidrule(lr){1-3}\cmidrule(lr){4-6}
            \textbf{V} & \textbf{A} &\multicolumn{1}{c}{\textbf{T}} & \textbf{ACC-2 (↑)} & \textbf{F1 (↑)} & \textbf{ACC-7 (↑)} \\
            \hline
            \ding{51} & \ding{55} & \ding{55} & 63.32 & 61.47 & 41.38 \\
            \ding{55} & \ding{51} & \ding{55} & 63.35 & 58.48 & 39.51 \\
            \ding{55} & \ding{55} & \ding{51} & 85.14 & 85.32 & 53.40 \\
            \hline
            \ding{51} & \ding{51} & \ding{55} & 65.24 & 63.13 & 39.60 \\
            \ding{51} & \ding{55} & \ding{51} & 86.24 & 86.21 & 53.42 \\
            \ding{55} & \ding{51} & \ding{51} & 86.98 & 86.78 & 54.13 \\
            \hline
            \ding{51} & \ding{51} & \ding{51} & \textbf{87.70} & \textbf{87.68} & \textbf{55.59} \\
            \hline
        \end{tabular}
    \centering
    \caption{Ablation studies for different multimodal fusion strategies on the CMU-MOSEI dataset.}
    \label{tab:ablation_for_modal}
\end{table}

\mysubsubsection{Impact of Different Modalities} Table~\ref{tab:ablation_for_modal} reports the performance of different modality combinations on CMU-MOSEI. Results show that adding more modalities consistently improves performance, with all three modalities combined yielding the best results. Among them, text contributes the most, while audio and video alone perform the worst. These trends align with previous findings~\cite{wang2023distribution, gao2024enhanced}. Notably, even with only Audio and Text, FINE achieves SOTA in ACC-2 and F1, demonstrating its effectiveness despite being designed for tri-modal input.

\mysubsubsection{Role of Key Components} 

\begin{table}
    \begin{center}
    \renewcommand{\arraystretch}{1.1} 
        \begin{tabular*}{0.9\linewidth}{c|*{3}{c}}
            \hline
            \textbf{Configs} & \textbf{ACC-2} & \textbf{F1} & \textbf{ACC-7} \\
            \hline
            w/o $\mathcal{L}_{UP}$ & 86.76 & 86.63 & 51.58 \\
            \hline
            \multicolumn{4}{c}{Role of \, " Mixture of Q-Formers "} \\
            \hline
            w/o MoQ & 86.96 & 86.90 & 42.93 \\
            w/o MoQ + TE & 86.21 & 86.01 & 50.76 \\
            w/o MoQ + TCN & 86.65 & 86.71 & 54.97 \\
            w/o MoQ + LSTM & 86.21 & 86.22 & 52.01 \\
            w/o MoQ + MLPs & 84.87 & 85.01 & 50.98 \\
            \hline
            \multicolumn{4}{c}{Role of \, " Factorized Task-Relevant Encoder "} \\
            \hline
            w/o $I_{str},I_{utr}$      & 86.79 & 86.64 & 54.95 \\
            w/o $I_{sha},I_{uni}$      & 86.32 & 86.40 & 52.80 \\
            w/o $I_{sha},I_{str}$      & 85.31 & 85.47 & 53.19 \\
            w/o $I_{uni},I_{utr}$      & 85.86 & 85.95 & 52.29 \\
            w/o $I_{sha}$              & 84.40 & 83.76 & 52.05 \\
            w/o FTRE & 86.16 & 85.82 & 53.12 \\
            w/o FTRE + MISA & 86.98 & 86.96 & 54.32 \\
            \hline
            \multicolumn{4}{c}{Role of \, " Dynamic Contrastive Queue "} \\
            \hline
            w/o DCQ & 86.41 & 86.31 & 48.42 \\
            w/o DCQ + ACCon & 86.76 & 86.64 & 51.43 \\
            w/o DCQ + SCL & 86.65 & 86.60 & 48.32 \\
            DCQ (SCL) & 86.85 & 86.90 & 50.30 \\
            \hline
            \textbf{FINE} & \textbf{87.70} & \textbf{87.68} & \textbf{55.59}  \\
            \hline
        \end{tabular*}
    \end{center}
    \caption{Ablation studies for the key components on the CMU-MOSEI dataset.}
    \label{tab:ablation_for_modules}
\end{table}

Table~\ref{tab:ablation_for_modules} presents the ablation results that verify the effectiveness of each key component in our proposed FINE framework.

MoQ plays an essential role in early-stage feature extraction. It dynamically selects from lightweight Q-Former experts to model fine-grained affective cues from multiple perspectives. Replacing MoQ with a standard Transformer Encoder (TE) or lightweight alternatives such as LSTM, TCN, or MLPs results in notable performance degradation. These baseline networks struggle to model token-level interactions effectively and are less robust to padding noise. Furthermore, their output structures are less compatible with the downstream FTRE module, leading to suboptimal integration and sentiment recognition.

FTRE is critical for disentangling modality-shared and modality-unique features while filtering irrelevant noise via task-aware constraints. Replacing FTRE with MISA~\cite{hazarika2020misa}, which performs modality-invariant and modality-specific decomposition, leads to significant drops in ACC-2 and F1. To further investigate the role of mutual information estimation, we ablate the task-aware constraint and observe that removing task relevance causes only marginal gains over existing methods. In contrast, preserving only task-aware components without disentanglement also fails to improve performance. Interestingly, even without task relevance, our MI-based decomposition still achieves SOTA-level results, suggesting that mitigating redundancy and enhancing complementarity alone is highly effective. When we isolate only shared or only unique branches, performance degrades drastically, confirming that the disentanglement of both is essential to balanced representation learning.

DCQ contributes by modeling long-term temporal dynamics and reducing incidental noise. Compared to standard Supervised Contrastive Learning (SCL)~\cite{khosla2020supervised}, DCQ consistently improves performance, owing to its class-wise sub-queue design and memory of historical task-relevant embeddings. Furthermore, incorporating ACCon into the contrastive loss further improves results, highlighting the advantage of modeling label-aware similarity in continuous sentiment spaces.

These ablation studies comprehensively validate the effectiveness and synergy of the FINE framework’s components. They emphasize the core value of feature disentanglement, and demonstrate that reducing redundancy, enhancing modality complementarity, and improving task relevance are all crucial—and interdependent—factors in achieving robust and expressive multimodal sentiment representations.

\section{Conclusion}

In this paper, we propose FINE, a multimodal sentiment analysis framework based on mutual information estimation. At its core, FINE employs a Factorized Task-Relevant Encoder to disentangle input features into shared and unique branches while suppressing task-irrelevant noise. This strategy reduces redundancy, enhances modality complementarity, and improves alignment with sentiment prediction. To support this process, we introduce a Mixture of Q-Formers for early fine-grained feature extraction and a Dynamic Contrastive Queue for long-term pattern modeling. Together, these modules enable FINE to construct robust and expressive representations under heterogeneous and asynchronous multimodal conditions. Experiments on multiple benchmarks demonstrate FINE’s superior performance and confirm the importance of disentanglement and task relevance in multimodal sentiment analysis.

\bibliography{main}

@article{soleymani2017survey,
    title={A survey of multimodal sentiment analysis},
    author={Soleymani, Mohammad and Garcia, David and Jou, Brendan and Schuller, Bj{\"o}rn and Chang, Shih-Fu and Pantic, Maja},
    journal={Image and Vision Computing},
    volume={65},
    pages={3--14},
    year={2017},
    publisher={Elsevier}
}

@inproceedings{melville2009sentiment,
    author = {Melville, Prem and Gryc, Wojciech and Lawrence, Richard D.},
    title = {Sentiment analysis of blogs by combining lexical knowledge with text classification},
    year = {2009},
    isbn = {9781605584959},
    publisher = {Association for Computing Machinery},
    address = {New York, NY, USA},
    IGNOREurl = {https://doi.org/10.1145/1557019.1557156},
    IGNOREdoi = {10.1145/1557019.1557156},
    booktitle = {Proceedings of the 15th ACM SIGKDD International Conference on Knowledge Discovery and Data Mining},
    pages = {1275–1284},
    numpages = {10},
    location = {Paris, France},
    series = {KDD '09}
}

@article{petrovica2017emotion,
    title={Emotion recognition in affective tutoring systems: Collection of ground-truth data},
    author={Petrovica, Sintija and Anohina-Naumeca, Alla and Ekenel, Haz{\i}m Kemal},
    journal={Procedia Computer Science},
    volume={104},
    pages={437--444},
    year={2017},
    publisher={Elsevier}
}

@article{liu2017facial,
    title={A facial expression emotion recognition based human-robot interaction system.},
    author={Liu, Zhentao and Wu, Min and Cao, Weihua and Chen, Luefeng and Xu, Jianping and Zhang, Ri and Zhou, Mengtian and Mao, Junwei},
    journal={IEEE CAA J. Autom. Sinica},
    volume={4},
    number={4},
    pages={668--676},
    year={2017}
}

@article{sanchez2019social,
    title={Social context in sentiment analysis: Formal definition, overview of current trends and framework for comparison},
    author={S{\'a}nchez-Rada, J Fernando and Iglesias, Carlos A},
    journal={Information Fusion},
    volume={52},
    pages={344--356},
    year={2019},
    publisher={Elsevier}
}

@INPROCEEDINGS{he2022masked,
    author={He, Kaiming and Chen, Xinlei and Xie, Saining and Li, Yanghao and Dollár, Piotr and Girshick, Ross},
    booktitle={2022 IEEE/CVF Conference on Computer Vision and Pattern Recognition (CVPR)}, 
    title={Masked Autoencoders Are Scalable Vision Learners}, 
    year={2022},
    volume={},
    number={},
    pages={15979-15988},
    IGNOREdoi={10.1109/CVPR52688.2022.01553}
}

@article{gandhi2023multimodal,
    title={Multimodal sentiment analysis: A systematic review of history, datasets, multimodal fusion methods, applications, challenges and future directions},
    author={Gandhi, Ankita and Adhvaryu, Kinjal and Poria, Soujanya and Cambria, Erik and Hussain, Amir},
    journal={Information Fusion},
    volume={91},
    pages={424--444},
    year={2023},
    publisher={Elsevier}
}

@INPROCEEDINGS{poria2016convolutional,
    author={Poria, Soujanya and Chaturvedi, Iti and Cambria, Erik and Hussain, Amir},
    booktitle={2016 IEEE 16th International Conference on Data Mining (ICDM)}, 
    title={Convolutional MKL Based Multimodal Emotion Recognition and Sentiment Analysis}, 
    year={2016},
    volume={},
    number={},
    pages={439-448},
    IGNOREdoi={10.1109/ICDM.2016.0055}
}

@inproceedings{morency2011towards,
    author = {Morency, Louis-Philippe and Mihalcea, Rada and Doshi, Payal},
    title = {Towards multimodal sentiment analysis: harvesting opinions from the web},
    year = {2011},
    isbn = {9781450306416},
    publisher = {Association for Computing Machinery},
    address = {New York, NY, USA},
    IGNOREurl = {https://doi.org/10.1145/2070481.2070509},
    IGNOREdoi = {10.1145/2070481.2070509},
    booktitle = {Proceedings of the 13th International Conference on Multimodal Interfaces},
    pages = {169–176},
    numpages = {8},
    location = {Alicante, Spain},
    series = {ICMI '11}
}

@article{nagrani2021attention,
    title={Attention bottlenecks for multimodal fusion},
    author={Nagrani, Arsha and Yang, Shan and Arnab, Anurag and Jansen, Aren and Schmid, Cordelia and Sun, Chen},
    journal={Advances in neural information processing systems},
    volume={34},
    pages={14200--14213},
    year={2021}
}

@ARTICLE{jiang2022sdn,
    author={Jiang, Xun and Xu, Xing and Zhang, Jingran and Shen, Fumin and Cao, Zuo and Shen, Heng Tao},
    journal={IEEE Transactions on Neural Networks and Learning Systems}, 
    title={SDN: Semantic Decoupling Network for Temporal Language Grounding}, 
    year={2024},
    volume={35},
    number={5},
    pages={6598-6612},
    IGNOREdoi={10.1109/TNNLS.2022.3211850}
}

@INPROCEEDINGS{fan2023pmr,
    author={Fan, Yunfeng and Xu, Wenchao and Wang, Haozhao and Wang, Junxiao and Guo, Song},
    booktitle={2023 IEEE/CVF Conference on Computer Vision and Pattern Recognition (CVPR)}, 
    title={PMR: Prototypical Modal Rebalance for Multimodal Learning}, 
    year={2023},
    volume={},
    number={},
    pages={20029-20038},
    IGNOREdoi={10.1109/CVPR52729.2023.01918}
}

@inproceedings{zhang2023provable,
    author = {Zhang, Qingyang and Wu, Haitao and Zhang, Changqing and Hu, Qinghua and Fu, Huazhu and Zhou, Joey Tianyi and Peng, Xi},
    title = {Provable dynamic fusion for low-quality multimodal data},
    year = {2023},
    publisher = {JMLR.org},
    booktitle = {Proceedings of the 40th International Conference on Machine Learning},
    articleno = {1753},
    numpages = {17},
    location = {Honolulu, Hawaii, USA},
    series = {ICML'23}
}

@article{mai2020modality, 
    title={Modality to Modality Translation: An Adversarial Representation Learning and Graph Fusion Network for Multimodal Fusion}, 
    volume={34}, 
    IGNOREurl={https://ojs.aaai.org/index.php/AAAI/article/view/5347}, 
    IGNOREdoi={10.1609/aaai.v34i01.5347}, 
    number={01}, 
    journal={Proceedings of the AAAI Conference on Artificial Intelligence}, 
    author={Mai, Sijie and Hu, Haifeng and Xing, Songlong}, 
    year={2020}, 
    month={Apr.}, 
    pages={164-172} 
}

@INPROCEEDINGS{tishby2015deep,
    author={Tishby, Naftali and Zaslavsky, Noga},
    booktitle={2015 IEEE Information Theory Workshop (ITW)}, 
    title={Deep learning and the information bottleneck principle}, 
    year={2015},
    volume={},
    number={},
    pages={1-5},
    IGNOREdoi={10.1109/ITW.2015.7133169}
}

@article{zhao2023fair, 
    title={Fair Representation Learning for Recommendation: A Mutual Information Perspective}, 
    volume={37}, 
    IGNOREurl={https://ojs.aaai.org/index.php/AAAI/article/view/25617}, 
    IGNOREdoi={10.1609/aaai.v37i4.25617}, 
    number={4}, 
    journal={Proceedings of the AAAI Conference on Artificial Intelligence}, 
    author={Zhao, Chen and Wu, Le and Shao, Pengyang and Zhang, Kun and Hong, Richang and Wang, Meng}, 
    year={2023}, 
    month={Jun.}, 
    pages={4911-4919} 
}

@InProceedings{su2023towards,
    author    = {Su, Weijie and Zhu, Xizhou and Tao, Chenxin and Lu, Lewei and Li, Bin and Huang, Gao and Qiao, Yu and Wang, Xiaogang and Zhou, Jie and Dai, Jifeng},
    title     = {Towards All-in-One Pre-Training via Maximizing Multi-Modal Mutual Information},
    booktitle = {Proceedings of the IEEE/CVF Conference on Computer Vision and Pattern Recognition (CVPR)},
    month     = {June},
    year      = {2023},
    pages     = {15888-15899}
}

@article{cha2022domain,
    title={Domain Generalization by Mutual-Information Regularization with Pre-trained Models},
    author={Junbum Cha and Kyungjae Lee and Sungrae Park and Sanghyuk Chun},
    journal={European Conference on Computer Vision (ECCV)},
    year={2022},
    pages={440--457}
}

@article{li2022invariant, 
    title={Invariant Information Bottleneck for Domain Generalization}, 
    volume={36}, 
    IGNOREurl={https://ojs.aaai.org/index.php/AAAI/article/view/20703}, 
    IGNOREdoi={10.1609/aaai.v36i7.20703}, 
    number={7}, 
    journal={Proceedings of the AAAI Conference on Artificial Intelligence}, 
    author={Li, Bo and Shen, Yifei and Wang, Yezhen and Zhu, Wenzhen and Reed, Colorado and Li, Dongsheng and Keutzer, Kurt and Zhao, Han}, 
    year={2022}, 
    month={Jun.}, 
    pages={7399-7407} 
}

@inproceedings{kurutach2018learning,
    author = {Kurutach, Thanard and Tamar, Aviv and Yang, Ge and Russell, Stuart and Abbeel, Pieter},
    title = {Learning plannable representations with causal InfoGAN},
    year = {2018},
    publisher = {Curran Associates Inc.},
    address = {Red Hook, NY, USA},
    booktitle = {Proceedings of the 32nd International Conference on Neural Information Processing Systems},
    pages = {8747–8758},
    numpages = {12},
    location = {Montr\'{e}al, Canada},
    series = {NIPS'18}
}

@article{seitzer2021causal,
    title={Causal influence detection for improving efficiency in reinforcement learning},
    author={Seitzer, Maximilian and Sch{\"o}lkopf, Bernhard and Martius, Georg},
    journal={Advances in Neural Information Processing Systems},
    volume={34},
    pages={22905--22918},
    year={2021}
}

@inproceedings{cheng2020club,
    author = {Cheng, Pengyu and Hao, Weituo and Dai, Shuyang and Liu, Jiachang and Gan, Zhe and Carin, Lawrence},
    title = {CLUB: a contrastive log-ratio upper bound of mutual information},
    year = {2020},
    publisher = {JMLR.org},
    booktitle = {Proceedings of the 37th International Conference on Machine Learning},
    articleno = {166},
    numpages = {10},
    series = {ICML'20}
}

@article{oord2018representation,
    title={Representation learning with contrastive predictive coding},
    author={Oord, Aaron van den and Li, Yazhe and Vinyals, Oriol},
    journal={arXiv preprint arXiv:1807.03748},
    year={2018}
}

@article{lepikhin2020gshard,
    title={Gshard: Scaling giant models with conditional computation and automatic sharding},
    author={Lepikhin, Dmitry and Lee, HyoukJoong and Xu, Yuanzhong and Chen, Dehao and Firat, Orhan and Huang, Yanping and Krikun, Maxim and Shazeer, Noam and Chen, Zhifeng},
    journal={arXiv preprint arXiv:2006.16668},
    year={2020},
    numpages={},
}

@article{fedus2022switch,
    title={Switch transformers: Scaling to trillion parameter models with simple and efficient sparsity},
    author={Fedus, William and Zoph, Barret and Shazeer, Noam},
    journal={Journal of Machine Learning Research},
    volume={23},
    pages={1--39},
    year={2022}
}

@inproceedings{dai2024deepseekmoe,
    title = "{D}eep{S}eek{M}o{E}: Towards Ultimate Expert Specialization in Mixture-of-Experts Language Models",
    author = "Dai, Damai  and
      Deng, Chengqi  and
      Zhao, Chenggang  and
      Xu, R.x.  and
      Gao, Huazuo  and
      Chen, Deli  and
      Li, Jiashi  and
      Zeng, Wangding  and
      Yu, Xingkai  and
      Wu, Y.  and
      Xie, Zhenda  and
      Li, Y.k.  and
      Huang, Panpan  and
      Luo, Fuli  and
      Ruan, Chong  and
      Sui, Zhifang  and
      Liang, Wenfeng",
    booktitle = "Proceedings of the 62nd Annual Meeting of the Association for Computational Linguistics (Volume 1: Long Papers)",
    month = aug,
    year = "2024",
    address = "Bangkok, Thailand",
    publisher = "Association for Computational Linguistics",
    IGNOREurl = "https://aclanthology.org/2024.acl-long.70/",
    IGNOREdoi = "10.18653/v1/2024.acl-long.70",
    pages = "1280--1297",
}

@inproceedings{gao2024enhanced,
    author = {Gao, Zixian and Hu, Disen and Jiang, Xun and Lu, Huimin and Shen, Heng Tao and Xu, Xing},
    title = {Enhanced Experts with Uncertainty-Aware Routing for Multimodal Sentiment Analysis},
    year = {2024},
    isbn = {9798400706868},
    publisher = {Association for Computing Machinery},
    address = {New York, NY, USA},
    IGNOREurl = {https://doi.org/10.1145/3664647.3680949},
    IGNOREdoi = {10.1145/3664647.3680949},
    booktitle = {Proceedings of the 32nd ACM International Conference on Multimedia},
    pages = {9650–9659},
    numpages = {10},
    location = {Melbourne VIC, Australia},
    series = {MM '24}
}

@inproceedings{li2023blip,
    author = {Li, Junnan and Li, Dongxu and Savarese, Silvio and Hoi, Steven},
    title = {BLIP-2: bootstrapping language-image pre-training with frozen image encoders and large language models},
    year = {2023},
    publisher = {JMLR.org},
    booktitle = {Proceedings of the 40th International Conference on Machine Learning},
    articleno = {814},
    numpages = {13},
    location = {Honolulu, Hawaii, USA},
    series = {ICML'23}
}

@inproceedings{zadeh2017tensor,
    title = "Tensor Fusion Network for Multimodal Sentiment Analysis",
    author = "Zadeh, Amir  and
      Chen, Minghai  and
      Poria, Soujanya  and
      Cambria, Erik  and
      Morency, Louis-Philippe",
    booktitle = "Proceedings of the 2017 Conference on Empirical Methods in Natural Language Processing",
    month = sep,
    year = "2017",
    address = "Copenhagen, Denmark",
    publisher = "Association for Computational Linguistics",
    IGNOREurl = "https://aclanthology.org/D17-1115/",
    IGNOREdoi = "10.18653/v1/D17-1115",
    pages = "1103--1114",
}

@inproceedings{tsai2019multimodal,
    title = "Multimodal Transformer for Unaligned Multimodal Language Sequences",
    author = "Tsai, Yao-Hung Hubert  and
      Bai, Shaojie  and
      Liang, Paul Pu  and
      Kolter, J. Zico  and
      Morency, Louis-Philippe  and
      Salakhutdinov, Ruslan",
    booktitle = "Proceedings of the 57th Annual Meeting of the Association for Computational Linguistics",
    month = jul,
    year = "2019",
    address = "Florence, Italy",
    publisher = "Association for Computational Linguistics",
    IGNOREurl = "https://aclanthology.org/P19-1656/",
    IGNOREdoi = "10.18653/v1/P19-1656",
    pages = "6558--6569",
}

@inproceedings{liang2024factorized,
     author = {Liang, Paul Pu and Deng, Zihao and Ma, Martin Q. and Zou, James Y and Morency, Louis-Philippe and Salakhutdinov, Ruslan},
     booktitle = {Advances in Neural Information Processing Systems},
     pages = {32971--32998},
     publisher = {Curran Associates, Inc.},
     title = {Factorized Contrastive Learning: Going Beyond Multi-view Redundancy},
     IGNOREurl = {https://proceedings.neurips.cc/paper_files/paper/2023/file/6818dcc65fdf3cbd4b05770fb957803e-Paper-Conference.pdf},
     volume = {36},
     year = {2023}
}

@inproceedings{zhao2025accon,
  title={ACCon: Angle-Compensated Contrastive Regularizer for Deep Regression},
  author={Zhao, Botao and Qu, Xiaoyang and Kang, Zuheng and Peng, Junqing and Xiao, Jing and Wang, Jianzong},
  booktitle={Proceedings of the AAAI Conference on Artificial Intelligence},
  volume={39},
  pages={22750--22758},
  year={2025}
}

@article{khosla2020supervised,
    title={Supervised contrastive learning},
    author={Khosla, Prannay and Teterwak, Piotr and Wang, Chen and Sarna, Aaron and Tian, Yonglong and Isola, Phillip and Maschinot, Aaron and Liu, Ce and Krishnan, Dilip},
    journal={Advances in neural information processing systems},
    volume={33},
    pages={18661--18673},
    year={2020}
}

@inproceedings{yang2023confede,
    title = "{C}on{FEDE}: Contrastive Feature Decomposition for Multimodal Sentiment Analysis",
    author = "Yang, Jiuding  and
      Yu, Yakun  and
      Niu, Di  and
      Guo, Weidong  and
      Xu, Yu",
    booktitle = "Proceedings of the 61st Annual Meeting of the Association for Computational Linguistics (Volume 1: Long Papers)",
    month = jul,
    year = "2023",
    address = "Toronto, Canada",
    publisher = "Association for Computational Linguistics",
    IGNOREurl = "https://aclanthology.org/2023.acl-long.421/",
    IGNOREdoi = "10.18653/v1/2023.acl-long.421",
    pages = "7617--7630",
}

@inproceedings{hu2022unimse,
    title = "{U}ni{MSE}: Towards Unified Multimodal Sentiment Analysis and Emotion Recognition",
    author = "Hu, Guimin  and
      Lin, Ting-En  and
      Zhao, Yi  and
      Lu, Guangming  and
      Wu, Yuchuan  and
      Li, Yongbin",
    booktitle = "Proceedings of the 2022 Conference on Empirical Methods in Natural Language Processing",
    month = dec,
    year = "2022",
    address = "Abu Dhabi, United Arab Emirates",
    publisher = "Association for Computational Linguistics",
    IGNOREurl = "https://aclanthology.org/2022.emnlp-main.534/",
    IGNOREdoi = "10.18653/v1/2022.emnlp-main.534",
    pages = "7837--7851",
}

@inproceedings{chen2020simple,
    author = {Chen, Ting and Kornblith, Simon and Norouzi, Mohammad and Hinton, Geoffrey},
    title = {A simple framework for contrastive learning of visual representations},
    year = {2020},
    publisher = {JMLR.org},
    booktitle = {Proceedings of the 37th International Conference on Machine Learning},
    articleno = {149},
    numpages = {11},
    series = {ICML'20}
}

@INPROCEEDINGS{he2020momentum,
    author={He, Kaiming and Fan, Haoqi and Wu, Yuxin and Xie, Saining and Girshick, Ross},
    booktitle={2020 IEEE/CVF Conference on Computer Vision and Pattern Recognition (CVPR)}, 
    title={Momentum Contrast for Unsupervised Visual Representation Learning}, 
    year={2020},
    volume={},
    number={},
    pages={9726-9735},
    IGNOREdoi={10.1109/CVPR42600.2020.00975}
}

@INPROCEEDINGS{caron2021emerging,
    author={Caron, Mathilde and Touvron, Hugo and Misra, Ishan and Jegou, Hervé and Mairal, Julien and Bojanowski, Piotr and Joulin, Armand},
    booktitle={2021 IEEE/CVF International Conference on Computer Vision (ICCV)}, 
    title={Emerging Properties in Self-Supervised Vision Transformers}, 
    year={2021},
    volume={},
    number={},
    pages={9630-9640},
    IGNOREdoi={10.1109/ICCV48922.2021.00951}
}

@InProceedings{zolfaghari2021crossclr,
    author    = {Zolfaghari, Mohammadreza and Zhu, Yi and Gehler, Peter and Brox, Thomas},
    title     = {CrossCLR: Cross-Modal Contrastive Learning for Multi-Modal Video Representations},
    booktitle = {Proceedings of the IEEE/CVF International Conference on Computer Vision (ICCV)},
    month     = {October},
    year      = {2021},
    pages     = {1450-1459}
}

@article{liang2022high,
    title={High-Modality Multimodal Transformer: Quantifying Modality \& Interaction Heterogeneity for High-Modality Representation Learning},
    author={Liang, Paul Pu and Lyu, Yiwei and Fan, Xiang and Tsaw, Jeffrey and Liu, Yudong and Mo, Shentong and Yogatama, Dani and Morency, Louis-Philippe and Salakhutdinov, Russ},
    journal={Transactions on Machine Learning Research},
    year={2022}
}

@inproceedings{sun2022cubemlp,
    author = {Sun, Hao and Wang, Hongyi and Liu, Jiaqing and Chen, Yen-Wei and Lin, Lanfen},
    title = {CubeMLP: An MLP-based Model for Multimodal Sentiment Analysis and Depression Estimation},
    year = {2022},
    isbn = {9781450392037},
    publisher = {Association for Computing Machinery},
    address = {New York, NY, USA},
    IGNOREurl = {https://doi.org/10.1145/3503161.3548025},
    IGNOREdoi = {10.1145/3503161.3548025},
    booktitle = {Proceedings of the 30th ACM International Conference on Multimedia},
    pages = {3722–3729},
    numpages = {8},
    location = {Lisboa, Portugal},
    series = {MM '22}
}

@inproceedings{rahman2020integrating,
    title = "Integrating Multimodal Information in Large Pretrained Transformers",
    author = "Rahman, Wasifur  and
      Hasan, Md Kamrul  and
      Lee, Sangwu  and
      Bagher Zadeh, AmirAli  and
      Mao, Chengfeng  and
      Morency, Louis-Philippe  and
      Hoque, Ehsan",
    booktitle = "Proceedings of the 58th Annual Meeting of the Association for Computational Linguistics",
    month = jul,
    year = "2020",
    address = "Online",
    publisher = "Association for Computational Linguistics",
    IGNOREurl = "https://aclanthology.org/2020.acl-main.214/",
    IGNOREdoi = "10.18653/v1/2020.acl-main.214",
    pages = "2359--2369",
}

@inproceedings{liu2018efficient,
    title = "Efficient Low-rank Multimodal Fusion With Modality-Specific Factors",
    author = "Liu, Zhun  and
      Shen, Ying  and
      Lakshminarasimhan, Varun Bharadhwaj  and
      Liang, Paul Pu  and
      Bagher Zadeh, AmirAli  and
      Morency, Louis-Philippe",
    booktitle = "Proceedings of the 56th Annual Meeting of the Association for Computational Linguistics (Volume 1: Long Papers)",
    month = jul,
    year = "2018",
    address = "Melbourne, Australia",
    publisher = "Association for Computational Linguistics",
    IGNOREurl = "https://aclanthology.org/P18-1209/",
    IGNOREdoi = "10.18653/v1/P18-1209",
    pages = "2247--2256",
}

@article{tsai2018learning,
    title={Learning factorized multimodal representations},
    author={Tsai, Yao-Hung Hubert and Liang, Paul Pu and Zadeh, Amir and Morency, Louis-Philippe and Salakhutdinov, Ruslan},
    journal = {International Conference on Representation Learning},
    year={2018}
}

@inproceedings{hazarika2020misa,
    author = {Hazarika, Devamanyu and Zimmermann, Roger and Poria, Soujanya},
    title = {MISA: Modality-Invariant and -Specific Representations for Multimodal Sentiment Analysis},
    year = {2020},
    isbn = {9781450379885},
    publisher = {Association for Computing Machinery},
    address = {New York, NY, USA},
    IGNOREurl = {https://doi.org/10.1145/3394171.3413678},
    IGNOREdoi = {10.1145/3394171.3413678},
    booktitle = {Proceedings of the 28th ACM International Conference on Multimedia},
    pages = {1122–1131},
    numpages = {10},
    location = {Seattle, WA, USA},
    series = {MM '20}
}

@inproceedings{han2021bi,
    author = {Han, Wei and Chen, Hui and Gelbukh, Alexander and Zadeh, Amir and Morency, Louis-philippe and Poria, Soujanya},
    title = {Bi-Bimodal Modality Fusion for Correlation-Controlled Multimodal Sentiment Analysis},
    year = {2021},
    isbn = {9781450384810},
    publisher = {Association for Computing Machinery},
    address = {New York, NY, USA},
    IGNOREurl = {https://doi.org/10.1145/3462244.3479919},
    IGNOREdoi = {10.1145/3462244.3479919},
    booktitle = {Proceedings of the 2021 International Conference on Multimodal Interaction},
    pages = {6–15},
    numpages = {10},
    location = {Montr\'{e}al, QC, Canada},
    series = {ICMI '21}
}

@INPROCEEDINGS{ma2023multimodal,
    author={Ma, Feipeng and Zhang, Yueyi and Sun, Xiaoyan},
    booktitle={2023 IEEE International Conference on Multimedia and Expo (ICME)}, 
    title={Multimodal Sentiment Analysis with Preferential Fusion and Distance-aware Contrastive Learning}, 
    year={2023},
    volume={},
    number={},
    pages={1367-1372},
    IGNOREdoi={10.1109/ICME55011.2023.00237}
}

@article{mai2022multimodal,
    title={Multimodal information bottleneck: Learning minimal sufficient unimodal and multimodal representations},
    author={Mai, Sijie and Zeng, Ying and Hu, Haifeng},
    journal={IEEE Transactions on Multimedia},
    volume={25},
    pages={4121--4134},
    year={2022},
    publisher={IEEE}
}

@ARTICLE{lin2023dynamically,
    author={Lin, Ronghao and Hu, Haifeng},
    journal={IEEE Transactions on Multimedia}, 
    title={Dynamically Shifting Multimodal Representations via Hybrid-Modal Attention for Multimodal Sentiment Analysis}, 
    year={2024},
    volume={26},
    number={},
    pages={2740-2755},
    IGNOREdoi={10.1109/TMM.2023.3303711}
}

@InProceedings{li2023decoupled,
    author    = {Li, Yong and Wang, Yuanzhi and Cui, Zhen},
    title     = {Decoupled Multimodal Distilling for Emotion Recognition},
    booktitle = {Proceedings of the IEEE/CVF Conference on Computer Vision and Pattern Recognition (CVPR)},
    month     = {June},
    year      = {2023},
    pages     = {6631-6640}
}

@INPROCEEDINGS{wang2023distribution,
    author={Wang, Yuanzhi and Cui, Zhen and Li, Yong},
    booktitle={2023 IEEE/CVF International Conference on Computer Vision (ICCV)}, 
    title={Distribution-Consistent Modal Recovering for Incomplete Multimodal Learning}, 
    year={2023},
    volume={},
    number={},
    pages={21968-21977},
    IGNOREdoi={10.1109/ICCV51070.2023.02013}
}

@inproceedings{vaswani2017attention,
    author = {Vaswani, Ashish and Shazeer, Noam and Parmar, Niki and Uszkoreit, Jakob and Jones, Llion and Gomez, Aidan N. and Kaiser, \L{}ukasz and Polosukhin, Illia},
    title = {Attention is all you need},
    year = {2017},
    isbn = {9781510860964},
    publisher = {Curran Associates Inc.},
    address = {Red Hook, NY, USA},
    booktitle = {Proceedings of the 31st International Conference on Neural Information Processing Systems},
    pages = {6000–6010},
    numpages = {11},
    location = {Long Beach, California, USA},
    series = {NIPS'17}
}

@article{zadeh2016mosi,
    title={MOSI: Multimodal Corpus of Sentiment Intensity and Subjectivity Analysis in Online Opinion Videos},
    author={Zadeh, Amir and Zellers, Rowan and Pincus, Eli and Morency, Louis-Philippe},
    journal={arXiv preprint arXiv:1606.06259},
    year={2016}
}

@inproceedings{zadeh2018multimodal,
    title = "Multimodal Language Analysis in the Wild: {CMU}-{MOSEI} Dataset and Interpretable Dynamic Fusion Graph",
    author = "Bagher Zadeh, AmirAli  and
      Liang, Paul Pu  and
      Poria, Soujanya  and
      Cambria, Erik  and
      Morency, Louis-Philippe",
    booktitle = "Proceedings of the 56th Annual Meeting of the Association for Computational Linguistics (Volume 1: Long Papers)",
    month = jul,
    year = "2018",
    address = "Melbourne, Australia",
    publisher = "Association for Computational Linguistics",
    IGNOREurl = "https://aclanthology.org/P18-1208/",
    IGNOREdoi = "10.18653/v1/P18-1208",
    pages = "2236--2246",
}

@inproceedings{devlin2019bert,
    title = "{BERT}: Pre-training of Deep Bidirectional Transformers for Language Understanding",
    author = "Devlin, Jacob  and
      Chang, Ming-Wei  and
      Lee, Kenton  and
      Toutanova, Kristina",
    booktitle = "Proceedings of the 2019 Conference of the North {A}merican Chapter of the Association for Computational Linguistics: Human Language Technologies, Volume 1 (Long and Short Papers)",
    month = jun,
    year = "2019",
    address = "Minneapolis, Minnesota",
    publisher = "Association for Computational Linguistics",
    IGNOREurl = "https://aclanthology.org/N19-1423/",
    IGNOREdoi = "10.18653/v1/N19-1423",
    pages = "4171--4186"
}

@inproceedings{baltruvsaitis2016openface,
    author={Baltrušaitis, Tadas and Robinson, Peter and Morency, Louis-Philippe},
    booktitle={2016 IEEE Winter Conference on Applications of Computer Vision (WACV)}, 
    title={OpenFace: An open source facial behavior analysis toolkit}, 
    year={2016},
    volume={},
    number={},
    pages={1-10},
    IGNOREdoi={10.1109/WACV.2016.7477553}
}

@INPROCEEDINGS{degottex2014covarep,
    author={Degottex, Gilles and Kane, John and Drugman, Thomas and Raitio, Tuomo and Scherer, Stefan},
    booktitle={2014 IEEE International Conference on Acoustics, Speech and Signal Processing (ICASSP)}, 
    title={COVAREP — A collaborative voice analysis repository for speech technologies}, 
    year={2014},
    volume={},
    number={},
    pages={960-964}
}

@article{kinney2014equitability,
  title={Equitability, mutual information, and the maximal information coefficient},
  author={Kinney, Justin B and Atwal, Gurinder S},
  journal={Proceedings of the National Academy of Sciences},
  volume={111},
  number={9},
  pages={3354--3359},
  year={2014},
  publisher={National Academy of Sciences}
}

@article{hasan2019ur,
  title={UR-FUNNY: A multimodal language dataset for understanding humor},
  author={Hasan, Md Kamrul and Rahman, Wasifur and Zadeh, Amir and Zhong, Jianyuan and Tanveer, Md Iftekhar and Morency, Louis-Philippe and others},
  journal={arXiv preprint arXiv:1904.06618},
  year={2019}
}

@inproceedings{yu2020ch,
  title={Ch-sims: A chinese multimodal sentiment analysis dataset with fine-grained annotation of modality},
  author={Yu, Wenmeng and Xu, Hua and Meng, Fanyang and Zhu, Yilin and Ma, Yixiao and Wu, Jiele and Zou, Jiyun and Yang, Kaicheng},
  booktitle={Proceedings of the 58th annual meeting of the association for computational linguistics},
  pages={3718--3727},
  year={2020}
}

@inproceedings{zadeh2018memory,
  title={Memory fusion network for multi-view sequential learning},
  author={Zadeh, Amir and Liang, Paul Pu and Mazumder, Navonil and Poria, Soujanya and Cambria, Erik and Morency, Louis-Philippe},
  booktitle={Proceedings of the AAAI conference on artificial intelligence},
  volume={32},
  year={2018}
}

@inproceedings{yu2021learning,
  title={Learning modality-specific representations with self-supervised multi-task learning for multimodal sentiment analysis},
  author={Yu, Wenmeng and Xu, Hua and Yuan, Ziqi and Wu, Jiele},
  booktitle={Proceedings of the AAAI conference on artificial intelligence},
  volume={35},
  pages={10790--10797},
  year={2021}
}

@article{li2024coupled,
  title={Coupled mamba: Enhanced multimodal fusion with coupled state space model},
  author={Li, Wenbing and Zhou, Hang and Yu, Junqing and Song, Zikai and Yang, Wei},
  journal={Advances in Neural Information Processing Systems},
  volume={37},
  pages={59808--59832},
  year={2024}
}

@inproceedings{pramanick2022multimodal,
  title={Multimodal learning using optimal transport for sarcasm and humor detection},
  author={Pramanick, Shraman and Roy, Aniket and Patel, Vishal M},
  booktitle={Proceedings of the IEEE/CVF winter conference on applications of computer vision},
  pages={3930--3940},
  year={2022}
}

\clearpage

\section{Appendix}

In this Appendix, we provide additional method details, experimental results, and further experimental details and findings.
\section{Related Work}

\mysubsubsection{Mutual Information} is a measure used in information theory to evaluate the dependence between two random variables~\cite{tishby2015deep}. Mutual information
(MI) stands out as an equitable measure that can capture relationships of any form, and generalizes across continuous, discrete, and multidimensional variables~\cite{kinney2014equitability}.  Mathematically, the Mutual Information between variables $x$ and $y$ s defined as:

\begin{equation}
    \begin{aligned}
    \mathrm{I}(X;Y) &= \, D_{KL}\left(p(x,y) || p(x)p(y)\right) 
    \\
    &= \, \mathbb{E}_{p(\boldsymbol{x},\boldsymbol{y})} \left[\log\frac{p(\boldsymbol{x},\boldsymbol{y})}{p(\boldsymbol{x})p(\boldsymbol{y})}\right] \, .
     \end{aligned}
\end{equation}

Previous studies have demonstrated the benefits of incorporating MI optimization into deep learning. It has proven to be effective in areas such as representation learning~\cite{su2023towards,zhao2023fair}, domain generalization~\cite{cha2022domain,li2022invariant}, and causality~\cite{kurutach2018learning,seitzer2021causal}. Direct estimation of MI in high-dimensional spaces is challenging; therefore, existing approaches typically rely on estimating the variational lower bounds~\cite{oord2018representation} and upper bounds ~\cite{cheng2020club} of MI. From a multimodal perspective, maximizing the lower bound of MI can be used to increase the correlation between modalities, while minimizing the upper bound of MI helps to reduce the correlation between modalities.

\mysubsubsection{Contrastive Learning}Contrastive learning is a widely used technique for learning discriminative feature representations by maximizing the similarity between positive pairs and minimizing that between negative pairs. This approach has been extensively explored in representation learning, with notable methods such as SimCLR~\cite{chen2020simple}, MoCO~\cite{he2020momentum}, DINO~\cite{caron2021emerging}, and CrossCLR~\cite{zolfaghari2021crossclr}. In the supervised setting, techniques like Supervised Contrastive Learning~\cite{khosla2020supervised} use class labels to define positive and negative pairs, achieving impressive performance in tasks like image classification.
We categorize samples within the same bin as positive pairs and those from different bins as negative pairs. Specifically, for an anchor \(x_i\), the positive pairs set is defined as \(P(i) := \{j \in B \mid y_i = y_j, j \neq i\}\), and the negative pairs set is defined as \(N(i) := \{j \in B \mid y_i \neq y_j\}\). A simple extension of the supervised contrastive learning loss function can be derived for regression tasks, as shown below:
\begin{equation}
    \mathcal{L}_i = - \frac{1}{|P(i)|} \sum_{p \in P(i)} \log \frac{\exp(\cos(\theta_{i,p})/\tau)}{\sum_{k \in N(i)\cup P(i)} \exp(\cos(\theta_{i,k})/\tau)},
\end{equation}
where \( \theta_{i,k} \) represents the angle between embeddings \( z_i \) and \( z_k \), and \( \cos(\theta_{i,k}) = z_i^T z_k \). However, a key challenge arises when contrastive learning is applied to continuous label classification or regression tasks. In these tasks, label distances are not simple categorical intervals but exhibit continuity and order. Existing contrastive methods often overlook these true distance relationships, leading to imprecise feature representations.

Furthermore, traditional contrastive learning methods often fail to effectively address issues such as class imbalance, sensitivity to incidental noise, and a lack of memory for long-term patterns. Although both Contrastive Feature Decomposition ~\cite{yang2023confede} and Towards Unified Multimodal Sentiment Analysis ~\cite{hu2022unimse} employ intra-modal and inter-modal contrastive learning, they still fail to effectively address the two challenges mentioned above. To overcome these limitations, we propose the Dynamic Contrastive Queue strategy, which maintains a dynamic queue of features and labels across time steps, allowing the model to learn long-term patterns while balancing memory across classes. Additionally, we integrate an Angle-Compensated Contrastive Regularizer to better capture the varying distances between labels in continuous tasks, ensuring more precise and meaningful feature representations.

\section{Methods}

\mysubsubsection{Mixture of Q-Formers} Mixture of Experts (MoE) is a technique for enhancing model representational capacity by expanding the parameter space. It typically comprises a gating network and multiple expert subnetworks, each specialized for a particular subspace of the data. A learnable gating mechanism selectively activates a subset of these experts based on the input, thereby dynamically adjusting the computation path and mitigating interference from noisy samples. In recent years, fueled by growing computational resources and data availability, MoE architectures have been widely adopted in large language models~\cite{lepikhin2020gshard,fedus2022switch,dai2024deepseekmoe}. MoE has also been successfully explored in multimodal sentiment analysis. For instance, enhanced experts with uncertainty-aware routing~\cite{gao2024enhanced} introduced an uncertainty-driven gating mechanism that enables the model to adaptively select experts based on input noise levels, leading to improved robustness and performance.

In this work, we propose a hybrid module called \textit{Mixture of Q-Formers (MoQ)}, which leverages a diverse set of experts to dynamically capture fine-grained and complementary features across modalities. In MoQ, each expert is instantiated using Q-Former~\cite{li2023blip}, which functions as an information bottleneck. Unlike conventional feedforward networks that process high-dimensional inputs directly, Q-Former employs learnable query tokens to extract compact representations. This design enables MoQ to retain key information while reducing computational overhead, thereby achieving a balance between fine-grained expressiveness and efficiency.

MoQ also incorporates a \textit{Gate Router}, which selects the most relevant subset of experts based on the input. Let \( Q \) and \( G \) denote the Q-Former and Gate Router, respectively. For a given modality feature \( X_m \in \mathbb{R}^{T_m \times d_m} \), where \( m \in \{T, A, V\} \) denotes the modality (Text, Audio, Video), \( T_m \) is the sequence length, and \( d_m \) is the feature dimensionality. The routing process is defined as:

\begin{align}
R(x_m^i) &= \mathrm{Linear}(\mathrm{AvgPool}(x_m^i)) \, , \\
G(x_m^i) &= \mathrm{Top}_k(\mathrm{Softmax}(R(x_m^i))) \, ,
\end{align}
where \( R(x_m^i) \in \mathbb{R}^N \) represents the unnormalized expert scores, followed by softmax normalization and top-$k$ selection.

The final output of MoQ for sample \( x_m^i \) is computed by aggregating the selected experts:

\begin{equation}
\hat{x}_m^i = \sum_{j=1}^{N} G_m^j(x_m^i) \cdot Q_m^j(x_m^i) \, ,
\end{equation}
where \( Q_m^j(x_m^i) \in \mathbb{R}^{l \times d_m'} \) is the output of the $j$-th expert, \( l \) is the number of output tokens, and \( d_m' \) is the hidden feature dimensionality. \( G_m^j(x_m^i) \in \mathbb{R}^{l \times d_m'} \) serves as the fusion weight for each expert output.

To prevent expert imbalance, we introduce an auxiliary load-balancing loss \( \mathcal{L}_{aux} \), inspired by prior work~\cite{lepikhin2020gshard,fedus2022switch}, to encourage uniform expert usage. The loss is defined as:

\begin{equation}
\mathcal{L}_{aux}= N \cdot \sum_{i=1}^{N} f^i_m \cdot P^i_m \, ,
\end{equation}
where \( f^i_m \) denotes the fraction of tokens assigned to expert \( i \):

\begin{equation}
f^i_m=\frac{1}{|\mathcal{B}|}\sum_{x_m\in \mathcal{B}}\mathbb{1}(\text{Token } x_m \text{ selects Expert } i) \, ,
\end{equation}
and \( P^i_m \) is the average routing probability of expert \( i \):

\begin{equation}
P^i_m=\frac{1}{|\mathcal{B}|}\sum_{x_m\in \mathcal{B}}G^i_m(x_m) \, .
\end{equation}

This auxiliary objective promotes more balanced expert participation, improving generalization and model stability.

\mysubsubsection{Dynamic Contrastive Queue}To mitigate the impact of spurious noise and capture more discriminative features, we propose a dynamic contrastive queue (DCQ) strategy. DCQ optimizes global features, allowing the model to capture long-term patterns beyond the local mini-batch context. Specifically, we construct a large queue \( \mathrm{Q} = (\mathrm{F}, \mathrm{L}) \) with \( K \) sub-queues, where \( \mathrm{F} \) represents the most recent feature embeddings and \( \mathrm{L} \) represents the most recent labels in \( \mathrm{Q} \). Each sub-queue retains the most recent task-relevant features across time steps, improving the model's memory and pattern recognition capabilities. Let \( N_i \) represent the sample size of class \( i \), \( \alpha \in (0, 1] \) be the adjustment factor, and \( S_{\text{min}} \) the minimum sub-queue length. The capacity \( S_i \) of the \( i \)-th sub-queue \( \mathrm{Q}_i^s \) is:
\begin{equation}
S_i = \max\left(\alpha \cdot N_i, S_{\text{min}}\right)
\end{equation}
where \( \alpha \) controls the impact of class size on the queue capacity, and \( S_{\text{min}} \) ensures a minimum capacity to preserve feature diversity. This approach ensures balanced feature storage across classes, with the minimum sub-queue length reducing the negative effects of extreme class size variations. When the more recent features and labels from the previous \(t-1\) steps have already been enqueued, and the queue contains the paired set \(\mathrm{F}^{t-1}\) and \(\mathrm{L}^{t-1}\), the queue is updated at step \(t\) as follows:
\begin{equation}
    (\mathrm{F}^t, \mathrm{L}^t) = \text{QueueUpdate} ( \mathrm{f}^{t}, \mathrm{l}^{t}), \{\mathrm{Q}_1^s, \mathrm{Q}_2^s, \dots, \mathrm{Q}_k^s\}^{t-1})  \, ,
\end{equation}
where $\mathrm{f}^t$ and $\mathrm{l}^t$ denote the normalized feature and label at time step \(t\), while $\text{QueueUpdate}$ represents the queue update operation.

We introduce the Angle-Compensated Contrastive Regularizer (ACCon) ~\cite{zhao2025accon}, a novel approach that adjusts the similarity between anchors and negative samples within the contrastive learning framework, overcoming the limitations of traditional methods in regression tasks. ACCon incorporates an angle compensation mechanism, enabling the model to adjust based on the actual distances between labels during optimization. Specifically, assuming a linear negative correlation between labels, we can compute the expected compensation angle $\varphi$ based on the distances between labels:
\begin{equation}
    \varphi = \pi\left( 1-\frac{y_{\mathrm{neg}}-y_{\mathrm{anc}}}{max(\mathcal{Y})-min(\mathcal{Y})} \right)  \, ,
\end{equation}
where \(y_{\mathrm{neg}}\) denotes the label of the negative sample, \(y_{\mathrm{anc}}\) refers to the label of the anchor sample, and \(\max(\mathcal{Y})\) and \(\min(\mathcal{Y})\) represent the maximum and minimum values in the label space \(\mathcal{Y}\), respectively. Consequently, the cosine similarity with angle compensation is expressed as:
\begin{equation}
    \begin{aligned}
        \cos(\tilde{\theta}_{i,m}) = & \cos(\hat{\theta}_{i,m} + \varphi)
        \\
         = & \, z_i z_m^T \cos(\varphi) - | \sin(\varphi) | \sqrt{1 - (z_i z_m^T)^2 + \epsilon}  \, ,
    \end{aligned}
\end{equation}
where \(\hat{\theta}_{i,m}\) is the original cosine angle between the anchor embedding \(z_i\) and the negative sample embedding \(z_m\), and \(\epsilon\) is the smoothing term. Finally, at step $t$, we can compute the angle-compensated contrastive loss\(\mathcal{L}^{i}_{CL}\) for sample $i$ with the queue:

\begin{equation}
    \mathcal{L}^{i}_{CL} = \frac{-1}{|\mathcal{P}(i)|}
    \sum_{p \in \mathcal{P}(i)} 
    \log \frac{\exp (z_{i} z_{p}^{T} / \tau)}
    {\left( 
    \substack{
            \sum_{k \in \mathcal{P}(i)} \exp (z_{i} z_{k}^{T} / \tau ) \\
            +\sum_{m \in \mathcal{N}(i)} \exp (\cos (\tilde{\theta}_{i, m}) / \tau)
    }
    \right)} \, ,
\end{equation}
Where $\mathcal{P}(i)$ and $\mathcal{N}(i)$ represent the positive set and negative set of anchor $i$ in $\mathrm{F}_{t-1} \cup \mathrm{f}_t$, respectively.

\begin{algorithm}
\caption{FINE Framework}
\label{alg:framework}
\KwIn{Dataset \(D\) with unimodal feature \(X_m, \, m \in \{T,V,A\} \), raw text \(X_T^r\) (if provided), multimodal labels \(y\), hyperparameters}
\KwOut{Prediction \(\hat{y}\)}
Initialize FINE\;
\While{not done}{
    \If{use BERT}{
        $X_T =BERT(X_T^r)$
    }
    Module: MoQ\;
    \For{modality $m \in \{T, A, V\}$}{
        $R(X_m) = \text{Linear}(\text{AvgPool}(X_{m}))$\;
        $G(X_m) = \text{Top}_{k}(\text{SoftMax}( R(X_m))) $\;
        $\hat{X}_m = \sum_{i=1}^N G_m^i(X_m) Q_m^i(X_m)$ \;
    }
    Module: FTRE\;
    \For{modality $m \in \{T, A, V\}$}{
        $X_{m}^s = E_m^{s}(\hat{X}_m)$\;
        $X_{m}^s = E_m^{u}(\hat{X}_m)$\;
        $X_{m}^{str} = E_m^{str}(X_{m}^s)$\;
        $X_{m}^{utr} = E_m^{utr}(X_{m}^u)$\;
    }
    $ X_T^{tr} = \text{ConCat}(X_{T}^{str},X_{T}^{utr}) $\;
    $ X_A^{tr} = \text{ConCat}(X_{A}^{str},X_{A}^{utr}) $\;
    $ X_V^{tr} = \text{ConCat}(X_{V}^{str},X_{V}^{utr}) $\;
    Module: Unimodal Decoder\;
    \For{modality $m \in \{T, A, V\}$}{
        $\hat{Q}_{m}^s = \text{Transformer Decoder}(X_m^{tr}, Q_m)$\;
        $\hat{y}_m = \text{MLP}(\hat{Q}_{m}^s$)\;
    }
    Module: Multimodal Decoder\;
    $ X^{tr} = \text{ConCat}(X_T^{tr},X_A^{tr},X_V^{tr}) $\;
    $ F^{tr} = \text{Transformer Encoder}(X^{tr})$\;
    $\hat{y} = \text{MLP}(\mathrm{F^{tr}})$\;
    Optimization object\;
  $\mathcal{L}_{total}= \mathcal{L}_{MP} + \lambda_{cl}\mathcal{L}_{CL} + \lambda_{up} \mathcal{L}_{UP} + \lambda_{aux}\mathcal{L}_{aux} +\beta_{mi} \mathcal{L}_{MI}$\;
}
\end{algorithm}

\mysubsubsection{FINE Framework Overview} In this paper, we present the FINE framework, a novel multimodal fusion method for sentiment analysis. The algorithm, detailed in Algorithm~\ref{alg:framework}, processes unimodal features, including text, audio, and visual inputs, through specialized modules such as MoQ, FTRE, and DCQ. The MoQ module enables dynamic feature fusion through modality-specific experts, while FTRE decouples shared and modality-specific information to enhance task-relevant features. DCQ improves long-term pattern learning by optimizing feature storage and reducing noise. 

\section{Experiments and Results}
\subsection{Dataset}

\mysubsubsection{CMU-MOSI \& CMU-MOSEI} CMU-MOSI consists of 2,199 short monologue video clips. The dataset is divided into 1,284 samples for training, 229 for validation, and 686 for testing. CMU-MOSEI contains 22,856 movie review video clips sourced from YouTube. According to the predefined protocol, 16,326 samples are used for training, while the remaining 1,871 and 4,659 samples are allocated for validation and testing, respectively. For both datasets, we extract linguistic features using a pre-trained BERT~\cite{devlin2019bert}, obtaining 768-dimensional hidden states as word representations. For the visual modality, we encode each video frame using Facet~\cite{baltruvsaitis2016openface}, which captures the presence of 35 facial action units. The acoustic modality was processed by COVAREP~\cite{degottex2014covarep} to obtain the 74-dimensional features. The sentiment labels for both datasets are manually annotated on a continuous scale ranging from -3 to 3, representing a spectrum of sentiment, including highly negative, negative, slightly negative, neutral, slightly positive, and highly positive. It is important to note that there is no overlap between the CMU-MOSI and CMU-MOSEI datasets, and their data collection and annotation processes are independent of each other.

\mysubsubsection{UR-FUNNY} is a multimodal humor detection dataset comprising 1,866 videos from 1,741 speakers and a total of 9,588 utterances. Following prior work, we use a cleaned version with overlapping and noisy instances removed and enriched with contextual information. Each utterance is paired with its surrounding context and annotated with a binary label indicating whether it is humorous or not. The dataset is split into 7,614 training samples, 980 validation samples, and 994 test samples. All utterances are sourced from TED talks, ensuring speaker and topic diversity. UR-FUNNY captures humor perception through multimodal signals, making it a valuable benchmark for multimodal humor understanding.

\mysubsubsection{CH-SIMS} is a high-quality Chinese multimodal sentiment analysis dataset containing 2,281 carefully curated video segments. Each sample is annotated with one multimodal sentiment label and three unimodal sentiment scores (text, audio, and visual), ranging from -1 (strongly negative) to +1 (strongly positive). To ensure real-life relevance, video clips were sourced from movies, TV shows, and variety programs. The original videos were manually trimmed at the frame level using Adobe Premiere Pro, a process that, while time-consuming, ensured high temporal precision. During data collection and preprocessing, the following constraints were applied: only Mandarin speech was considered, clips with accented speech were handled cautiously, and each segment was between 1 and 10 seconds long. Additionally, only one visible speaker face was allowed per clip to maintain clarity. CH-SIMS features a diverse set of speakers across various age groups, making it a robust and realistic benchmark for multimodal sentiment understanding in Chinese.

\begin{table*}[!ht]
    \centering
    \begin{tabular}{lcccc}
        \toprule
        \textbf{Hyperparameter} & \textbf{CMU-MOSI} & \textbf{CMU-MOSEI}& \textbf{UR-FUNNY} & \textbf{CH-SIMS}\\
        \midrule
        batch size & 32 & 64& 128 & 32 \\
        learning rate & 6e-5 & 3e-5 & 3e-5 & 5e-5\\
        BERT learning rate & 4e-5 & 3e-5 & 1e-5 & 3e-5 \\
        warmup epochs & \multicolumn{4}{c}{0.1} \\
        total epochs & 100 & 50 & 100 & 50 \\
        num of experts & 4 & 8 & 8 & 8 \\
        num of query tokens & 4 & 8 & 8 & 8\\
        top-k ratio & \multicolumn{4}{c}{0.75} \\
        MoQ dim \((a,t,v)\) & (128,256,256) & (256,256,256)  & (256,256,256) &  (256,512,512)\\
        reduction ratio & \multicolumn{4}{c}{0.5} \\
        \(\mathcal{\alpha}\) & 0.7 & 0.3 & 0.3 & 0.5 \\
        \(\mathcal{\lambda}_{CL}\) & 1.0 & 3.0 & 1.0 & 1.0 \\
        \(\mathcal{\lambda}_{UP}\) & 0.4 & 0.5& 0.5 & 0.5 \\
        \(\mathcal{\lambda}_{aux}\) & 0.2 & 1.0 & 0.2 & 0.2 \\
        \(\mathcal{\beta}_{mi}\) & 0.5 & 0.5& 0.3 & 0.3 \\
        \bottomrule
    \end{tabular}
    \caption{Training Hyperparameters in each dataset.}
    \label{tab:hyper}
\end{table*}

\subsection{Evaluation Metrics}
\mysubsubsection{CMU-MOSI \& CMU-MOSEI} Consistent with previous studies~\cite{li2023decoupled, gao2024enhanced}, we assess FINE’s performance using the following standard metrics: (1) Binary Accuracy (Acc-2): This measures the accuracy for the binary classification task; (2) F1 Score: The F1 score is calculated for each category to provide a balance between precision and recall;  (3) Accuracy-7 (Acc-7): This metric measures the accuracy across seven different sentiment intensity categories. For the CMU-MOSI and CMU-MOSEI datasets, Acc-2 and F1 scores are reported in the non-excluding zero form: negative/positive.
\mysubsubsection{UR-FUNNY} The task is a standard binary classification with binary accuracy (Acc-2) as the metric for evaluation.
\mysubsubsection{CH-SIMS} We adopt three widely used metrics: Binary Accuracy (Acc-2),  F1 Score, and  Accuracy-7 (Acc-7).

\subsection{Baselines}
In this experiment, we selected recent state-of-the-art multimodal models as baselines for a comprehensive comparison. These include models such as TFN~\cite{zadeh2017tensor}, LMF~\cite{liu2018efficient}, and GFN~\cite{mai2020modality}, which directly fuse multimodal features. MISA~\cite{hazarika2020misa}, DMD~\cite{li2023decoupled} and ConFEDE~\cite{yang2023confede} focus on the importance of modality-specific tokens. MulT~\cite{tsai2019multimodal} explores a novel attention mechanism. 
C-MFN~\cite{hasan2019ur} is an extension of the Memory Fusion Network~\cite{zadeh2018memory}, which introduces a memory-augmented neural architecture that explicitly models cross-modal interactions and their temporal evolution for multi-view sequential learning.
MAGBERT~\cite{rahman2020integrating} introduces a Multimodal Adaptation Gate to adapt pre-trained language models like BERT and XLNet for multimodal sentiment analysis by injecting vision and acoustic cues as learned shifts in their internal representations during fine-tuning.
Self-MM~\cite{yu2021learning}  leverages self-supervised label generation to create unimodal supervisions, and jointly trains multimodal and unimodal tasks with a dynamic weighting strategy to learn both consistency and difference among modalities.
CubeMLP~\cite{sun2022cubemlp} and C-MIB~\cite{mai2022multimodal}, which uses mutual information for denoising, also serve as key benchmarks. BBFN~\cite{han2021bi} and MSG~\cite{lin2023dynamically} investigate pairwise modality combinations while maintaining modality independence. Finally, EUAR~\cite{gao2024enhanced} considers the impact of noisy data on the internal features of each individual modality.

\subsection{Implementation Details}All experiments were conducted using the PyTorch framework on a GTX 3090 GPU with 24GB of memory, CUDA 12.1 and PyTorch version 2.3.0. For training, we employed the AdamW optimizer and used a fixed set of random seeds \{3585, 7154, 8757\} to ensure reproducibility. In all datasets, we consistently utilized 8 experts during the training process. More details can be found in the Supplementary Material. Hyperparameter settings on different datasets are shown in the Table~\ref{tab:hyper}, where 'reduction ratio' refers to the dimensionality reduction ratio of the encoder in FTRE.

\subsection{Complexity Analysis}

In this section, we measure FINE's time and space complexity. Taking CMU-MOSI as a representative case, we measure training time over 100 epochs under identical conditions and define space as the total number of parameters. FINE introduces approximately 15\% more parameters than the average across baselines, primarily attributed to the inclusion of Q-Formers and MLPs used for mutual information estimation. However, these components are excluded during inference, and thus do not impact deployment cost. Importantly, FINE achieves the fastest runtime among all compared models by avoiding sequential structures such as TCN or LSTM, and instead leveraging Transformers for efficient parallelism and hardware acceleration.

\begin{table}[ht]
\centering
\begin{tabular}{c|c|c}
\toprule
\textbf{Model} & \textbf{Params}~(↓) & \textbf{Running Time}~(↓) \\
\midrule
MAGBERT & 110,705,665 & 549s \\
MISA & 110,620,273 & 535s \\
C-MIB & 109,835,748 & 480s \\
TFN & 161,409,399 & 275s \\
ConFEDE & 129,601,154 & 194s \\
EUAR & 125,599,814 & 415s \\
\textbf{FINE} & 140,976,482 & 135s \\
\bottomrule
\end{tabular}
\caption{Comparison of parameter count and training time on CMU-MOSI.}
\label{tab:efficiency}
\end{table}

\subsection{Ablation Studies}

\begin{figure}[htbp]
    \centering
    \includegraphics[width=0.43\textwidth]{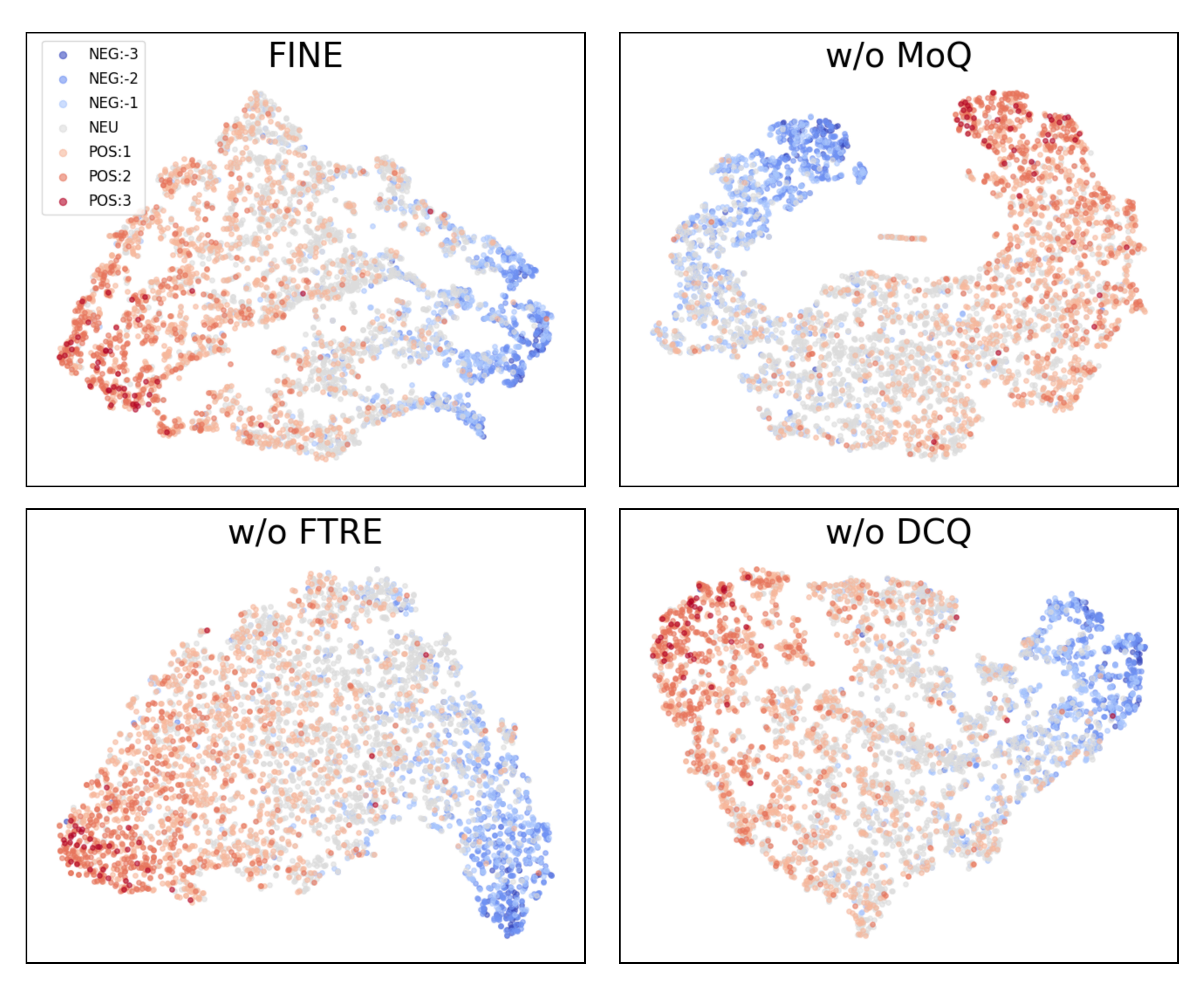}
    \caption{Visualization of ablation experiments for key components.}
    \label{tsne}
\end{figure}

\mysubsubsection{Visualizating Representations} We conducted a seven-class ablation experiment visualization on the CMU-MOSEI dataset, as shown in Figure~\ref{tsne}. We used the last layer of features before the classification head. In the visualization, the deeper the blue, the higher the degree of negativity, and the deeper the red, the higher the degree of positivity. The top-left plot in Figure~\ref{tsne} shows the feature visualization of the FINE method, which includes the complete model with all modules integrated. The other three plots show ablation experiments, demonstrating the visualization results after removing key components such as MoQ, FTRE, and DCQ.

The visualization in the top-left plot (FINE) clearly shows improved separation between the negative (blue) and positive (red) classes, with minimal misclassifications in the middle region, where weakly positive and weakly negative samples lie. This distribution fits well with the hypothesis that a greater difference between labels should result in a greater separation in their respective feature representations. The plot for the FINE model supports this hypothesis, as the separation between labels is more distinct. In contrast, the visualization of the experimental results without FTRE violates this hypothesis, as the distance between the classes is not positively correlated. The plot without FTRE shows a more scattered distribution, suggesting that the model fails to capture discriminative representations from different modalities effectively. Similarly, the visualization without DCQ lacks a sharp distinction between positive and negative classes, indicating that DCQ plays a key role in enhancing the separation of these classes.

\begin{figure}[htbp]
    \centering
    \includegraphics[width=0.40\textwidth]{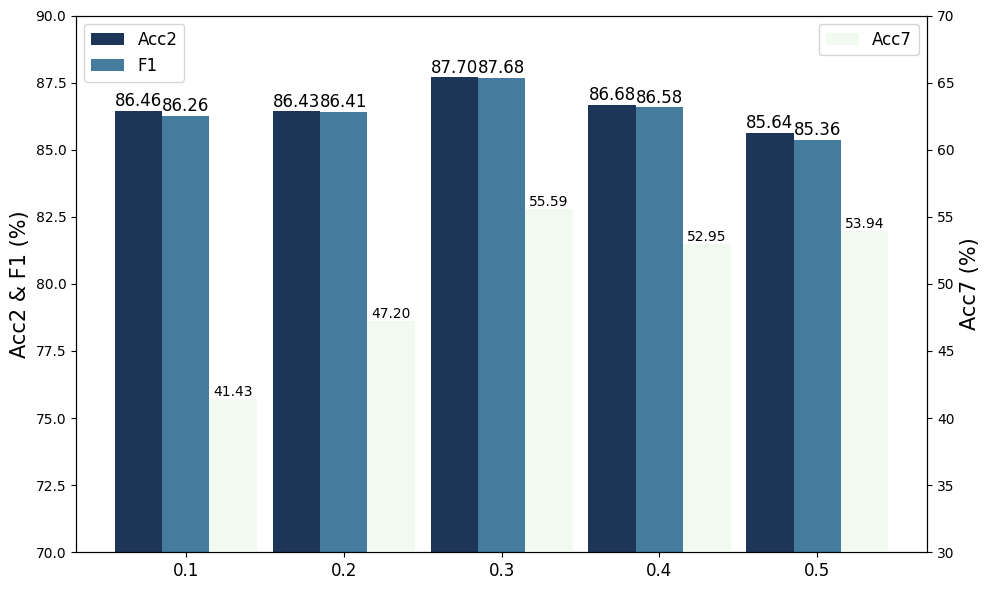}
    \caption{The impact of the adjustment factor $\alpha$ in DCQ on the MOSEI experimental results.}
    \label{fig:histogram}
\end{figure}

\begin{figure}[htbp]
    \centering
    \includegraphics[width=0.40\textwidth]{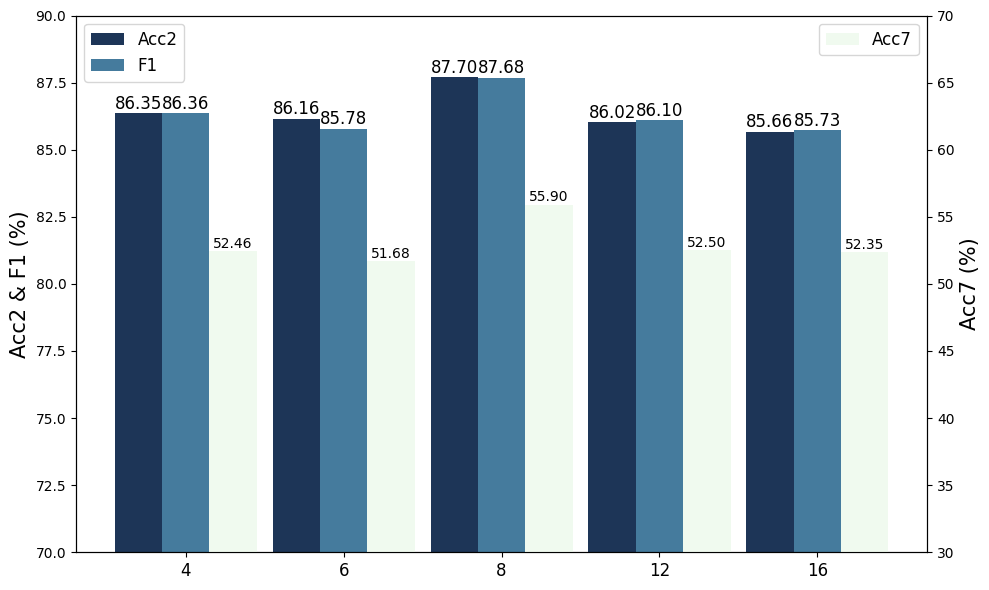}
    \caption{The impact of the experts' num $N$ in MoQ on the MOSEI experimental results.}
    \label{fig:histogram2}
\end{figure}
Then, we present additional ablation experiments that explore the impact of two hyperparameters on the experimental results: the adjustment factor \(\alpha\) in the DCQ module and the number of experts \(N\) in the MoQ module. Figure~\ref{fig:histogram} illustrates the influence of the adjustment factor \(\alpha\) on the performance of the model, where the results show that \(\alpha = 0.3\) yields the best performance across different evaluation metrics. Similarly, Figure~\ref{fig:histogram2} demonstrates the effect of varying the number of experts \(N\) in MoQ, with \(N = 8\) showing the most favorable results. Based on these findings, we conclude that \(\alpha = 0.3\) and \(N = 8\) represent the optimal settings for our experiments.

\end{document}